 \documentclass[10pt,journal]{IEEEtran}

\usepackage{amsmath,url} 
\usepackage{amssymb}  

\usepackage{graphicx,subfigure,cite,bm,lipsum}
\usepackage{multicol}
\graphicspath{ {images/} }
\usepackage{amsmath}
\usepackage{bbm}
\usepackage{widetext}
\usepackage{complexity}

\newcommand{\bP}[1]{{\mathbb{P}}\left[{#1}\right]}
\newcommand{\bE}[1]{{\mathbb{E}}\left[{#1}\right]}

\newcommand{\1}[1]{{\bf 1}\left[#1\right]}
\usepackage{mathtools}
\usepackage{amssymb}
\newtheorem{theorem}{Theorem}[section]

\title{Modeling and Analysis of Cascading Failures in Interdependent Cyber-Physical Systems}

\author{Yingrui Zhang and Osman Ya\u{g}an
\thanks{This research was supported by National Science Foundation through grant CCF \#1422165.}
\thanks{Parts of the material will be presented at  57th IEEE International Conference on Decision and Control (CDC 2018), Miami Beach (FL), Dec. 2018.}
\thanks{Y. Zhang and O. Ya\u{g}an are with the Department
of Electrical and Computer Engineering,
Carnegie Mellon University, Pittsburgh,
PA, 15213 USA. Email:
{\tt \small yingruiz@andrew.cmu.edu, oyagan@ece.cmu.edu}}}

\begin{document}

\maketitle

\begin{abstract}
Integrated cyber-physical systems (CPSs), such as the smart-grid, are increasingly becoming the underpinning technology for major industries. A major concern regarding such systems are the seemingly unexpected large scale failures, 
which are often attributed to a small initial shock getting escalated due to intricate dependencies 
within and across the 
individual (e.g., cyber and physical) counterparts  of the system. 
In this paper,
 we develop a novel interdependent system model to capture this phenomenon, also known as cascading failures. Our framework consists of two networks that have inherently different characteristics governing their {\em intra}-dependency: i) a {\em cyber}-network where a node is deemed to be functional as long as it belongs to the largest connected (i.e., giant) component; and ii) a {\em physical} network where nodes are given an initial {\em flow} and a {\em capacity}, and failure of a node results with redistribution of its flow to the remaining nodes, upon which further failures might take place due to {\em overloading} (i.e., the flow of a node exceeding its capacity). 
 Furthermore, it is assumed that these two networks are {\em inter-dependent}. For simplicity, we consider a one-to-one interdependency model where every node in the cyber-network is dependent upon and supports a single node in the physical network, and vice versa.
  We provide a thorough analysis of the dynamics of cascading failures in this interdependent system initiated with a random attack. The system robustness is quantified as the {\em surviving} fraction of nodes at the end of cascading failures, and  is derived in terms of 
all network parameters involved (e.g., degree distribution, load/capacity distribution, failure size, etc.). Analytic results are  supported through an extensive numerical study. Among other things, these results demonstrate the ability of our model to capture the {\em unexpected} nature of large-scale failures, and provide insights on improving system robustness.
\end{abstract}

\maketitle


%
\maketitle

\section{Introduction}\label{sec:introduction}

Today's worldwide network infrastructure consists of a web of
interacting cyber-networks (e.g., the Internet) and physical
systems (e.g., the power grid). Integrated cyber-physical systems (CPSs) are increasingly becoming the underpinning
technology for major industries.
The smart grid is an archetypal example of a CPS where the
power grid network and the communication network for its
operational control are coupled together; the grid depends on the communication
network for its control, and the communication network depends on the grid for power.
While this coupling with a communication network brings unprecedented improvements and functionality to the power grid, it has
been observed \cite{Vespignani} that such interdependent systems tend
to be fragile against failures, natural hazards, and attacks.
For instance, in the event of an attack or random failures in an interdependent
system, the failures in one of the networks can cause failures of
the dependent nodes in the other network and vice versa. This
process may continue in a recursive manner, triggering a
cascade of failures that can potentially collapse
an entire system. In fact, the cascading effect of even a
partial Internet blackout could disrupt major national
infrastructure networks involving Internet services, power grids,
and financial markets \cite{buldyrev2010catastrophic}. For example, it was shown \cite{Rosato} that 
the electrical blackout
that affected much of Italy on 28 September 2003 had
started with the shutdown of a power station, which led to failures in the Internet
communication network, which in turn caused the breakdown of more stations, and 
so on.

With interdependent systems becoming an integral part of our daily lives,
a fundamental question arises as to how we can design an interdependent system in a {\em robust} manner.
Towards this end, a major focus has to be put on understanding their vulnerabilities, and in particular the root cause of 
the seemingly unexpected but large scale cascading failures. 
These events are often attributed to a small initial shock getting escalated due to the intricate dependencies within and across the 
individual (e.g., cyber and physical) counterparts  of the system. 
Therefore, a good understanding of the robustness of many 
real-worlds systems passes through an accurate
characterization and modeling of these inherent dependencies. 

Traditional studies in network science fall short in characterizing the robustness of interdependent networks since the
focus has mainly been on single networks in isolation; i.e., networks that do not 
interact with, or depend on any other network.   
Despite some recent research activity aimed at studying interdependent networks  \cite{buldyrev2010catastrophic,BuldyrevShereCwilich,parshani1,HuangGaoBuldyrevHavlinStanley,ZhouGao_Partial_Scale_Free,DongGaoDu}, very few
consider engineering aspects of inter-dependent networks and very little is known as to how such systems can be designed to have maximum
robustness under certain design constraints; see  \cite{YaganQianZhangCochranLong,zhang2016optimizing,SundaramJSAC,chattopadhyay} for rare exceptions. 
The current literature is also lacking 
interdependent system models that capture fundamental differences between {\em physical} and {\em cyber} networks, and enable studying robustness of systems that integrate networks with inherently different behavior. For example, it would be expected that the functionality of the physical subsystem is primarily governed by the physical flows and capacities associated with its components, whereas  system-wide connectivity would be the prominent requirement for maintaining functionality in the cyber network.
There is thus a need to develop
 new approaches for modeling and analyzing cascading failures in interdependent cyber-physical systems.

In this paper, we develop a  model that will help understand how failures would propagate in an interdependent system that  constitutes physical {\em and} cyber networks. This requires characterization of {\em intra}-dependency models for each constituent network as well as an {\em inter}-dependency model describing the spread of failures {\em across} networks; see Section \ref{subsec:intra_vs_inter} for details. 
As already mentioned, the main drawback of the current literature on interdependent networks is that the focus has almost exclusively been on {\em percolation}-based failure models, where a node can function only if it belongs to the largest  connected (i.e., giant) component in the networks. While suitable for cyber or communication networks, such  models are not appropriate for networks carrying physical flows; 
e.g., in  power grid, {\em islanding} is a commonly used strategy for preventing cascades \cite{d2017curtailing}.

Our interdependent system model consists of  i)  a cyber-network where a node is assumed to be
functional as long as it belongs to the largest connected (i.e., giant) component; and ii) a {\em physical} network where nodes are given an initial {\em flow} and a {\em capacity}, and failure of a node results with redistribution of its flow to the remaining nodes, upon which further failures might take place due to {\em overloading} (i.e., the flow of a node exceeding its capacity). 
For simplicity, we consider a one-to-one interdependency model where every node in the cyber-network is dependent upon and supports a single node in the physical network, and vice versa.
Thus, a node in the cyber-network (resp.~physical network) will continue to function if and only if its support in the physical network (resp.~cyber-network) is functional {\em and} it belongs to the largest connected subgraph of the cyber-network (resp.~its capacity is larger than its current flow); see Section \ref{sec:model} for a detailed description of the system model.

We provide a thorough analysis of the dynamics of cascading failures in this interdependent system, where failures are initiated by a {\em random} attack on a certain fraction of nodes. The system robustness, defined as the {\em steady-state} fraction of nodes that survive the cascade, is characterized in terms of 
all network parameters involved (e.g., degree distribution of the cyber-network, load-capacity values in the physical-network, network size, attack size, etc.). Analytic results are supported by an extensive numerical study. 
An interesting finding is that under our model, the system goes through a {\em complete breakdown} through a {\em discontinuous} 
transition with respect to increasing attack size. In other words, 
 the variation of the \lq\lq mean fraction of functional nodes at the steady state" with respect to \lq\lq attack size" has a discontinuity at the {\em critical} attack size above which the system collapses. This indicates that our model's behavior is 
 reminiscent of large but rare blackouts seen in real world, and thus might help explain how small initial shocks can cascade to disrupt large systems that have proven stable with respect to similar disturbances in the past.

We also leverage our main result to investigate how the robustness can be improved by adjusting various parameters defining the interdependent system; e.g., load/capacity values in the physical network and the degree distribution of the cyber-network.  This can prove useful in designing an interdependent system so that it has maximum robustness under given constraints. It is important to note that limited prior work revealed  unprecedented differences in the behaviors of interdependent networks as compared to single networks. For instance, it has been shown \cite{buldyrev2010catastrophic,YaganQianZhangCochranLong} that a network design that is optimal in countering node failures in a single network could be the most catastrophic choice for the resiliency of interdependent networks.
For the model considered here, our results reveal an intricate connection between the robustness of each  constituent network when they are isolated and the robustness of the interdependent system formed by them. First of all, when all else is fixed, and the {\em total} capacity available to all nodes in the physical network is given, the interdependent system becomes  more robust when capacities are allocated such that every node has the same {\em redundant space} (i.e., capacity minus initial load) as compared to the commonly used \cite{MotterLai,WangChen,Mirzasoleiman,CrucittiLatora} allocation where  nodes  are given a redundant space proportional to their initial load. 
However, the situation becomes much more intricate when the degree distribution of the cyber-network and the redundant space allocation in the physical network are adjusted simultaneously. There, we observe that depending on the degree distribution of the cyber-network, an interdependent system with equal redundant space allocation can be more or less robust than one where redundant space is proportional to load (with mean node degree and initial loads fixed). Also, in contrast with the well-known results in single networks \cite{BarabasiAlbert} where degree distributions with large variance (e.g., Pareto) are associated with higher robustness (against {\em random} failures) than cases where the variance is small (e.g., Poisson distribution), we demonstrate that the comparison is more intricate for interdependent systems. In particular, we provide several examples where the interdependent system with a Pareto-distributed cyber-network is more or less robust than one where the cyber-network has Poisson degree distribution, even when all other parameters are kept constant.


We believe this work brings a new perspective to the field of robustness of interdependent networks 
and might help steer the literature away from the heavily-studied percolation models towards flow-redistribution models, {\em and}  towards models that combine networks with inherently different cascade characteristics (of which CPS is an archetypal example); to the best of our knowledge, this is the first work where the interdependence of two networks with fundamentally different cascade behavior is studied. We  believe that our results provide interesting insights on the robustness of interdependent CPSs against random failures and attacks. In particular, despite the simplicity of the models used, our results might capture  the {\em qualitative} behavior of cascades in an interdependent system well. We  also believe this work  will trigger further studies (and provide initial ideas) on how node capacities in the physical-network and the topology of the communication network can be designed jointly to maximize the robustness of an interdependent CPS. 


The rest of the paper is organized as follows. In Section \ref{sec:model}, we present our interdependent system model in details, starting with the 
distinction between intra-dependency and inter-dependency. 
In Section \ref{sec:results}, we present the main result of the paper, which allows computing the fraction of surviving nodes  at each step of cascading failures initiated by a random attack. Here, we also provide an outline of the proof, while full proof is given in Appendix.
In Section \ref{sec:numerical}, we present numerical results demonstrating the accuracy of our analysis in the finite node regime.  The paper is concluded in Section \ref{sec:conclusion} with several suggestions for future work.

\section{System Model}
\label{sec:model}
\subsection{Intra-dependency vs.~Inter-dependency}
\label{subsec:intra_vs_inter}
Our modeling framework is motivated with the inherent dependencies that exist in 
many real-world systems including
  cyber-physical systems (CPSs).
 Namely, we will characterize how component failures
propagate and cascade, both within the cyber or the physical parts of the system (due to \lq\lq intra-dependency"), as well as across them due to \lq\lq inter-dependency". The actual meaning of  \lq\lq failure" is expected to be domain-dependent and can vary from a component being physically damaged to a node's inability to carry out its tasks.
For ease of exposition, we consider two sub-systems, say $A$ and $B$. 

Assume that network $A$ consists of nodes $\{a_1, \ldots, a_N\}$ and network $B$ consists of nodes $\{b_1, \ldots, b_N\}$. For illustration purposes, we can think of network $A$ as the power network consisting of generators and substations (i.e., the physical network), and network $B$ as the control and communication network consisting of control centers and routers (i.e., the cyber network) -- This is a classical example of an interdependent CPS, with the power stations sending data to and receiving control signals from routers, and routers receiving power from substations.
Modeling the dependencies within and between networks $A$ and $B$ amounts to answering three questions. First, for both networks we have to decide on the set of rules governing how failures would propagate within that network, leading to a characterization of the {\em intra}-dependencies. For example, we should identify how the failure of a power node $a_i$ affects other substations and generators in the power network $A$. 
Similarly, we should identify how the failure of a communication node $b_j$ affects other 
nodes in $B$. Finally, we must characterize the {\em inter}-dependence of the two systems, and how this interdependence may lead to propagation of failures across them. Namely, we must have a set of rules that specify how the failure of a power station $a_i$ impacts the nodes $\{b_1, \ldots, b_N\}$ in the communication network and vice versa.

 \begin{figure}[!t]
\centering
{\includegraphics[width=0.34\textwidth]{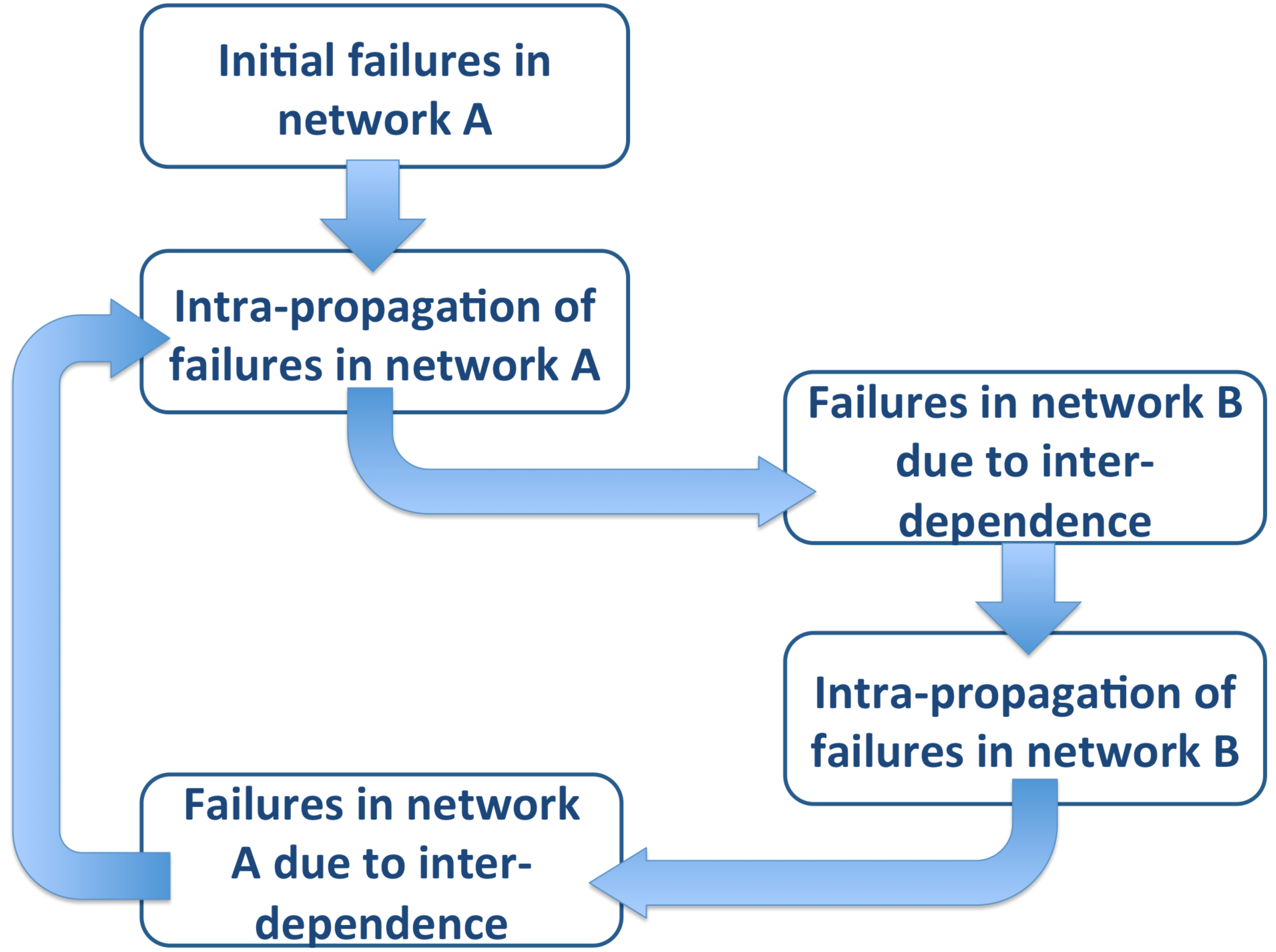}}
\caption{\sl An illustration of failure propagation model in an interdependent system.}
\label{fig:failure_process}
\vspace{-3mm}
\end{figure}

Once these modeling questions are answered, 
the propagation of failures in an interdependent system (consisting of networks $A$ and $B$) can be studied. Without loss of generality,  assume that the failures are
initiated in network $A$ due to random failures or attacks. To get a better idea about the role of intra- and inter-dependencies in the cascade of failures, consider an {\em asynchronous} failure update model, where the effect of intra-dependencies and inter-dependencies are considered in two separate batches, following one another. 
See Figure \ref{fig:failure_process} for an illustration of the asynchronous failure propagation model. The asynchronous failure update assumption eases the implementation and analysis of the model, and can be shown to yield the same steady-state network structures with a synchronous failure update model; just note that failure propagation process is monotone and that (according to our assumption) nodes can not heal once failed.

\subsection{The Model}
\label{subsec:detailed_model}
Despite the vast literature on interdependent networks \cite{buldyrev2010catastrophic,YaganQianZhangCochranLong,shahrivar2017spectral,ZhangArenasYagan}, there has been little (if any) attempt to characterize the robustness of  interdependent systems where the constituent networks have different intra-dependency behaviors. In the case of CPS, it would be expected that the cyber and physical counterparts obey inherently different rules governing how failures would propagate internally in each network.
 To this end, we study in this paper an interdependent system model that consists of two networks with   different characteristics governing their {\em intra}-dependency: i) a {\em cyber}-network where a node is deemed to be functional as long as it belongs to the largest connected (i.e., giant) component; and ii) a {\em physical} network where nodes are given an initial {\em flow} and a {\em capacity}, and failure of a node results with redistribution of its flow to the remaining nodes, upon which further failures might take place due to {\em overloading} (i.e., the flow of a node exceeding its capacity). To the best of our knowledge, this is the first work in the literature that studies interdependence between networks with fundamentally different intra-dependency; most existing works are focused on the interdependency between two physical networks (that obey a flow-redistribution-based model) \cite{scala2016cascades}, or two cyber-networks (that obey a giant-component-based intra-failure model) \cite{buldyrev2010catastrophic}.

\begin{figure}[!t]
\centering
{\includegraphics[width=0.5\textwidth]{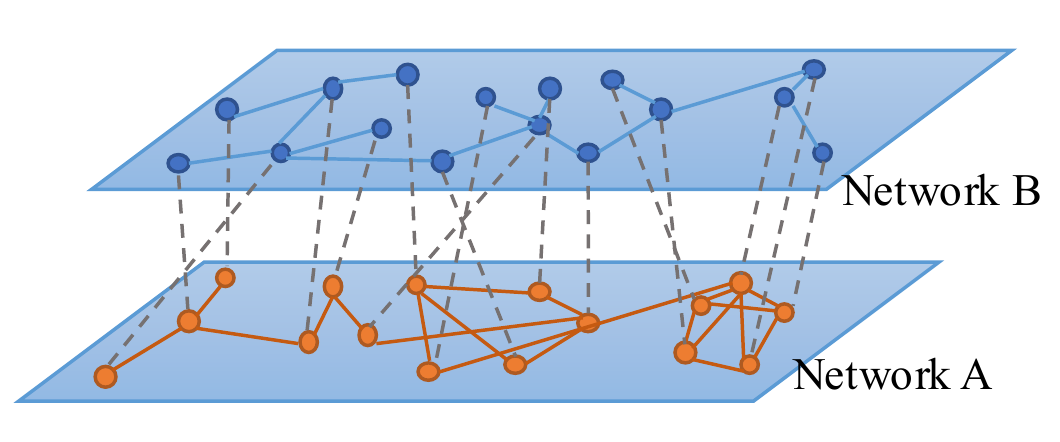}}
\caption{\sl System model illustration for the cyber-physical systems, where network $A$ can be the physical grid, and network $B$ can be the communication network that sends control signals. The interdependence across the two networks are realized through random one-to-one support links shown  by dashed lines.}
\label{fig:system_model}
\end{figure}

 For simplicity, the interdependence across the two networks is assumed to be one-to-one; i.e., every node in the cyber-network is dependent upon and supports a single node in the physical network, and vice versa; see Figure \ref{fig:system_model}. More precisely, we assume that for each $i=1,\ldots,N$, nodes $a_i$ and $b_i$ are dependent on each other meaning that if one fails, the other will fail as well. Although simplistic, the one-to-one interdependence model is considered to be a good starting point and has already provided useful insights in similar settings \cite{buldyrev2010catastrophic}; more complicated interdependence models shall be considered in future work including regular allocation strategy, i.e., each node in $A$ is connected to $k$ nodes in $B$ and vice versa, or a more general case where some nodes do not have interdependent links and can function even without any support from the other network.

\vspace{2mm}
\noindent
{\bf Intra-dependency in Network $A$.} Let network $A$ represent a flow network on nodes $a_1,\ldots,a_N$. Each node $a_i$ is given 
an initial load (e.g., power flow) $L_1, \ldots, L_N$.
The {\em capacity} $C_i$  of  node $a_i$ defines the maximum flow that it can sustain, and 
is  given by 
\begin{equation}
C_i= L_i + S_i, \qquad i=1,\ldots, N,
\label{eq:capacity}
\end{equation}
where $S_i$ denotes the {\em free-space} (or, redundancy) available to node $a_i$. 
It is assumed that a node {\em fails} (i.e., outages) if its load exceeds its capacity at any given time. The key assumption of our intra-dependency model for network $A$ is that when a node fails, the load it was carrying (right before the failure)
is redistributed {\em equally} among all remaining nodes. 
This leads to an increase in load carried by all remaining nodes, which in turn may lead to further failures of overloaded nodes, and so on, potentially leading to a cascade of failures.

The equal load redistribution rule takes its roots from the {\em democratic} fiber bundle model \cite{andersen1997tricritical,daniels1945statistical},
and has been recently used by Pahwa et al. \cite{pahwa2014abruptness} in the context  of power systems;
see also \cite{yagan_fiber_bundle,zhang2016optimizing}.
The relevance of the equal load-redistribution model for power systems stems from its ability to capture the {\em long-range} nature of the Kirchhoff's law, at least in the mean-field sense, as opposed to the {\em topological} models where failed load is  redistributed only {\em locally} among neighboring lines \cite{CrucittiLatora,WangRong}. 

Throughout we assume that the load and free-space pairs  $(L_i, S_i)$ are independently and identically distributed with $P_{LS}(x,y):=\bP{L \leq x, S \leq y}$ for each $i=1,\ldots, N$. The corresponding (joint) probability density
function is given by $p_{LS}(x,y) = \frac{\partial^2}{\partial x \partial y} P_{LS}(x,y)$.
 In order to avoid trivial cases, we assume that $S_{i}>0$ and $L_{i}>0$ with probability one for each $a_i$. Finally, we assume that the marginal densities $p_{L}(x)$ and $p_{S}(y)$ are continuous on their support.


\vspace{2mm}
\noindent
{\bf Intra-dependency in Network $B$.} 
Let network $B$ represent a cyber (e.g., communication) network consisting of nodes $b_1, \ldots, b_N$.
In this network, we assume that a node keeps functioning as long as it belongs to the largest (i.e., {\em giant}) connected
component of the network. 
If a node loses its connection to the giant core of the network, then it is assumed to have failed and can no longer carry out its functions. This percolation-based failure rule,
though not suitable for {\em physical} systems carrying a flow, 
can be regarded as a reasonable model for {\em cyber}-networks (e.g., sensor networks) where connectivity to a giant core would be crucial for a node's capability to deliver its tasks.

Robustness of networks under the giant-component based failure model has been extensively analyzed in the case of {\em single} networks \cite{BarabasiAlbert,newman2002spread,newman2001random}.
The focus has recently been shifted
towards {\em interdependent} networks with the work of Buldyrev et al.
\cite{buldyrev2010catastrophic}, where robustness of two interdependent networks, both operating under the giant-component based intra-dependence rule,
was studied. Their model, and most works that follow, are unable to capture the true nature of a cyber-physical network, where the cyber-network and the physical-network should obey a different set of rules determining their intra-dependencies. 

We define the structure of the network ${B}$
through its {\em degree distribution}, namely the
  probabilities $\{d_k, \: k=0,1,\ldots\}$ that
an arbitrary node in ${B}$ has degree $k$; clearly, we need to have 
$\sum_{k=0}^{\infty} d_k = 1$. 
In particular, each node $b_1,\ldots, b_N$ 
is assigned a degree drawn from the distribution $\{d_k\}_{k=0}^{\infty}$
independently from any other node. Once the degree sequence, $\textrm{degree}(b_1), \ldots, \textrm{degree}(b_N)$, of the network is determined, network $B$ is constructed by selecting {\em uniformly
at random} a graph among all graphs on $N$ nodes with the given degree sequence;
see \cite{MolloyReed,Bollobas,newman2001random} for details of such constructions. This class of networks is known in the literature as the {\em configuration model} or {\em random graphs with arbitrary degree distribution}. Degree distribution is often regarded 
as the core property defining a graph, and random networks with arbitrary degree distributions are extensively
used as a starting point in the literature on robustness of complex networks.

\vspace{2mm}
\noindent
{\bf Interdependent System Model.}
With the intra-dependency models of both networks specified, we adopt a one-to-one inter-dependency model across networks $A$ and $B$; i.e., nodes $a_i$ and $b_i$ depend on each other for each $i=1,\ldots, N$. With these in mind, we are interested in understanding the dynamics of cascading failures in this interdependent system, where failures are initiated by removing a $(1-p)$-fraction of nodes, selected  {\em randomly}, from network $A$ . As explained in Figure \ref{fig:failure_process}, we assume an asynchronous cascade model, where intra-propagation
and inter-propagation of failures are considered in a sequential manner. At any stage $t=1, 2, \ldots$ of the cascade process, a 
node $a_i$ in network $A$ will still be functioning if and only if (i) its current flow at time $t$ is less than its capacity; {\em and} (ii) its counterpart $b_i$ in network $B$ is still functioning (which is equivalent to $b_i$ being contained in the largest connected subgraph of $B$). Similarly, a node $b_j$ in network $B$ survives cascade step $t$ if and only if i) it belongs to largest connected component of $B$ at time $t$; and (ii) its counterpart $a_j$ in network $A$ is still functioning (which is equivalent to $a_j$ carrying a flow at time $t$ that is less than its capacity). 

Since the cascade process is monotone, a steady-state will eventually be reached, possibly after all nodes have failed. Let $\mathcal{N}_{\textrm{surviving}} \subset \{1, \ldots, N\}$ be the set of node id's that are still functioning at the steady state. In other words, the surviving interdependent system will consist of nodes $\{a_i: i \in \mathcal{N}_{\textrm{surviving}}\}$ where each $a_i$ has more capacity than its flow and 
$\{b_i: i \in \mathcal{N}_{\textrm{surviving}}\}$ that constitutes a connected subgraph of network $B$. The primary goal of this paper is to derive the {\em mean} fraction of nodes that survive the cascades as a function of the initial attack size $1-p$, in the asymptotic limit of large network size $N$. More precisely, we would like to characterize $S(p)$ defined as
\[
S(p) := \lim_{N \to \infty} \frac{\bE{|\mathcal{N}_{\textrm{surviving}}(p)|}}{N}
\]

In what follows, we present our main result that allows computing $S(p)$ under any degree distribution $\{d_k\}_{k=0}^{\infty}$ of the cyber-network $B$, any load-(free-space) distribution $P_{LS}(x,y)$ of the physical network $A$, and under any attack size $0 \leq p \leq 1$. This is followed in Section \ref{sec:numerical} by a numerical study that demonstrates the accuracy of our analysis even with finite $N$, and presents insights on how the robustness of an interdependent cyber-physical system can be improved by careful allocation of available resources (e.g., node capacities and degrees).



\section{Main Result}
\label{sec:results}

Our main result is presented next. The approach is based on recursively deriving the {\em mean} fraction of surviving nodes from both networks at each stage $t=1, 2, \ldots$ of the cascade process. The cascade process starts at time $t=0$ with a random attack that kills $1-p$ fraction of the nodes from network $A$. As mentioned earlier, we assume an asynchronous cascading failure model where at stages $t=1, 3, \ldots$ we consider the failures in network $A$ and in stages $t=2, 4, \ldots$ we consider the failures in network $B$. In this manner, we keep track of the subset of vertices $A_1 \supset A_3 \supset \ldots \supset A_{2i+1}$ and
$B_2 \supset B_4 \supset \ldots \supset B_{2i}$ that represent the functioning (i.e., surviving) nodes at the corresponding stage of the cascade. We let $f_{A_{i}}$ denote the {\em relative} size of the surviving set of nodes from network $A$ at stage $i$, i.e.,
\[
f_{A_{i}} = \frac{|A_i|}{N}, \quad i = 1, 3, 5, \ldots
\]
We define $f_{B_i}$ similarly as
\[
f_{B_{i}} = \frac{|B_i|}{N}, \quad i = 2, 4, 6, \ldots
\]
Our main result, presented next, shows how these quantities can be computed in a recursive manner.

\begin{theorem}
\label{thm:main}
{\sl Consider an interdependent system as described in Section \ref{sec:model},
where the load and free-space values of nodes $a_1, \ldots, a_N$ are drawn independently from the distribution $p_{LS}$, and network $B$ is generated according to the configuration model with degree distribution $\{d_k\}_{k=0}^{\infty}$; i.e., we have $\bP{\textrm{degree of node $b_i$} = k} = d_k$ for each $k=0,1, \ldots$ and $i=1,\ldots, N$.
Let mean degree be denoted by $\langle d \rangle$, i.e., let $\langle d \rangle=\sum_{k=0}^{\infty} k d_{k}$.
With
$f_{B_0}=p_{B_0}=p$, $f_{A_{-1}}=1$, and $Q_{-1}=0$, the relative size of the surviving parts of network $A$ and $B$ at each stage of the cascade, initiated by a random attack on $1-p$ fraction of the nodes, can be computed recursively as follows for each $i=0,1, \ldots$
\begin{widetext}
\begin{align}
   & p_{A_{2i+1}}   = \frac{f_{B_{2i}}}{f_{A_{2i-1}}} \label{eq:recurisive_osy_p_A} \\
 & Q_{2i+1}  = Q_{2i \hspace{-.3mm}-1} \hspace{-.5mm}+ \hspace{-.3mm}   \min  \hspace{-.5mm}\left\{ \hspace{-.3mm}x  \hspace{-.5mm}\in \hspace{-.5mm} (0, \infty] \hspace{-.3mm}: \hspace{-.5mm} \frac{\bP{S > Q_{2i-1}  \hspace{-.5mm}+ \hspace{-.5mm}  x}}{ \hspace{-.5mm}\bP{S > Q_{2i-1}}}(x  \hspace{-.5mm}+ \hspace{-.3mm} Q_{2i-1}  \hspace{-.4mm}+ \hspace{-.3mm} \bE{L ~|~ S  \hspace{-.5mm}> \hspace{-.3mm} x  \hspace{-.3mm}+\hspace{-.3mm} Q_{2i \hspace{-.3mm}-1}})  \hspace{-.5mm}\geq \hspace{-.3mm} \frac{Q_{2i-1}  \hspace{-.5mm}+ \hspace{-.3mm} \bE{L~|~S \hspace{-.3mm}> \hspace{-.3mm} Q_{2i-1}}}{p_{A_{2i+1}}}  \hspace{-.5mm}\right\} \label{eq:recurisive_osy_q}\\ 
    & f_{A_{2i+1}}  = f_{A_{2i-1}}\cdot p_{A_{2i+1}}\cdot \bP{S>Q_{2i+1} ~|~ S > Q_{2i-1}} \label{eq:recurisive_osy_f_A} \\
    & \nonumber\\
   & p_{B_{2i+2}} = p_{B_{2i}} \frac{f_{A_{2i+1}}}{f_{B_{2i}}} \label{eq:recurisive_osy_p_B} \\
   & u_{2i+2}  = \max \left\{u \in [0,1]: u= 1-\sum_{k=0}^{\infty} \frac{k d_k}{\langle d \rangle}  (1-u \cdot p_{B_{2i+2}})^{k-1}  \right\} \label{eq:recurisive_osy_u} \\ 
   & f_{B_{2i+2}} = p_{B_{2i+2}} \left( 1 - \sum_{k=0}^{\infty} d_k \left(1- u_{2i+2}\cdot p_{B_{2i+2}}\right)^k \right) \label{eq:recurisive_osy_f_B}
\end{align}
\end{widetext}
}
\end{theorem}
The notation used in Theorem \ref{thm:main} is summarized in Table \ref{tb:notation}.
In these iterations, it is assumed that if at any stage $i$, it happens to be the case that no $x<\infty$ satisfies the inequality at (\ref{eq:recurisive_osy_q}), we set $Q_{2i+1} =\infty$. It is then understood that the entire network $A$ (and thus $B$) have failed, and we get $f_{A_{2i+1}} = f_{B_{2i+2}} = 0$. Similarly, it can be seen that the equality in (\ref{eq:recurisive_osy_u}) always holds with $u=0$. Thus, if at any stage $i$, there is no $u>0$ satisfying the equality in (\ref{eq:recurisive_osy_u}), we will get $u_{2i+2}=0$ leading to $f_{B_{2i+2}}=0$; i.e., the entire network $B$ (and thus $A$) will have collapsed. 

\begin{table}[!h]
\begin{center}
\vspace{-1mm}
\begin{tabular}{|l|l| }
\hline  $A_i$ &  set of surviving nodes in network $A$ at stage $i=1, 3, 5, \ldots$\\
\hline  $B_i$ &  set of surviving nodes in network $B$ at stage $i=2, 4, 6, \ldots$\\
\hline
$f_{A_{2i+1}}$ &  fraction $|A_{2i+1}|/N$ of surviving nodes in  $A$ at stage $2i+1$\\
\hline
$f_{B_{2i+2}}$ &  fraction $|B_{2i+2}|/N$ of surviving nodes in  $B$ at stage $2i+2$\\
\hline
$Q_{2i+1}$ &  extra load per surviving node in $A$ at stage $2i+1$\\
\hline
$p_{A_{2i+1}}\hspace{-1mm}$ &  prob.~of a node in $A_{2i-1}$ surviving  {\em inter}-failures at stage $2i\hspace{-2mm}$ \\
\hline
$1-p_{B_{2i+2}}\hspace{-1mm}$ & 
{\em equivalent} prob.~of random attack to $B$ that gives ${B_{2i+2}}$\\
\hline
$u_{2i+2}$ &  auxiliary variable used in computing $f_{B_{2i+2}}\hspace{-1mm}$\\
\hline
\end{tabular}
\end{center}
 \caption{Key notation  in the analysis of cascading
failures}
\vspace{-5mm}
\label{tb:notation}
\end{table}

As mentioned before, our goal is to obtain the {\em final} system size, i.e., the relative size of the surviving nodes at the steady-state. In view of the one-to-one interdependence model, the surviving size of the networks $A$ and $B$ will be the same at the steady-state. Thus, 
we conclude that 
\[
S(p) = \lim_{i \to \infty} f_{A_{i}} = \lim_{i \to \infty} f_{B_{i}}.
\]

Next, we provide an outline of the proof, while the full details are available in Appendix. In \cite{zhang2016optimizing}, we already analyzed the cascade dynamics and derived the final system size
in a single flow carrying network (similar to network $A$ in our analysis),     
 when $1-p$ fraction of its nodes are randomly removed;
 the result enables computing the final system size in terms of the initial attack size $1-p$, as well as the load and free space distribution $P_{LS}(x,y)$.
 The results established in \cite{zhang2016optimizing} are incorporated in the recursions above through expression (\ref{eq:recurisive_osy_q}) that allows us to calculate, in a recursive manner, the extra load that each of the surviving nodes at a particular stage will be carrying in addition to their initial load.

 According to the failure propagation model described at the beginning of this section, at odd stages failures from network $B$ can propagate to network $A$, causing a fraction of nodes to be removed. As explained in details in Appendix, given that the intra-failure dynamics of network $B$ is completely {independent} from network $A$, the impact of the failures in $B$ to network $A$ will be  equivalent to a {\em random} attack launched on $A$. 
In addition,  at each odd stage $t=2i-1$, $i=1,2,\cdots$, we can treat the remaining part of network $A$ as a new physical network $A_{2i-1}$, 
 with the appropriately updated 
  size and load-\lq free-space' distribution.
  Thus, the random removal of nodes caused by failures in network $B$ (through the one-to-one interdependency links) from last cascade stage can be viewed as a new random attack to $A_{2i-1}$ that keeps only $p_{A_{2i+1}}$ fraction of its nodes alive. Then following a similar approach, we can compute the size of network $A$ at the next stage $2i+1$, i.e., $f_{A_{2i+1}}$. An important observation is the need to update the load and free-space distributions for each new network $A_{2i+1}$ to incorporate the facts that the surviving nodes in $A_{2i+1}$ are added with $Q_{2i-1}$ amount of extra load, and at the same time the free-space of each surviving node must be at least  $Q_{2i-1}$. We show in the detailed proof in Appendix that the changes of the distribution can be represented by the initial load and free-space distribution with $Q_{2i+1}$ representing the extra load in each stage. In other words, each time failures propagate between the two networks, network $A$ will shrink to a group of nodes that have a higher free space and that are now carrying more load. The fractional size of this surviving subset of nodes at each time stage can be computed via the equivalent attack size $p_{A_{2i+1}}$ (caused by failures in network $B$ propagated via the one-to-one dependent links), extra load $Q_{2i+1}$ and the load free-space distribution $P_{LS}(x,y)$; see (\ref{eq:recurisive_osy_p_A})-(\ref{eq:recurisive_osy_f_A}).

Following the same approach, in network $B$ we treat each new failure that comes from network $A$ as a new random attack (or failure) on the existing network $B_{2i+2}$. For a node in network $B$ to function, it must belong to the largest connected (i.e., giant) component, so actually the functioning network $B_{2i+2}$ at time stage $t=2i+2,i=0,1,2,\cdots$ is the giant component after the random attack propagated from network $A$. A key insight here is that the sequential process of applying a first random attack on the cyber-network, then computing the giant component, and then applying a second random attack and then computing the giant component is {\em equivalent} to (in terms of the fractional size of the set of nodes that survives) the process where the second random attack is applied directly after the first one without computing the giant component; e.g., see \cite{buldyrev2010catastrophic}. This way, the result of a series of random attack/giant-component calculation processes can be emulated by a single random attack/giant-component calculation, with an appropriately calculated {\em equivalent} random attack size.  In our calculations, this 
{\em equivalent} attack size for stage $2i+2$ is represented by $1-p_{B_{2i+2}}$ and can be computed recursively as given in (\ref{eq:recurisive_osy_p_B}). 
This formula is based on treating 
all {\em new} failures propagated from network $A$ in the following time stage as the new random attack size launched on $B$, which is then used to update the equivalent attack size $1-p_{B_{2i+2}}$ that will be used to emulate the entire cascade sequence up until that stage. 
Then, the size of network $B_{2i+2}$, namely the size of the giant component after randomly removing $(1-p_{B_{2i+2}})$-fraction of nodes, can be computed using the technique of generating functions \cite{newman2002spread,buldyrev2010catastrophic,gao2012networks,newman2001random,shao2011cascade}. 
 The formulas that give the network size $f_{B_{2i+2}}$ at each time stage $i=0,1, \ldots$ are presented at (\ref{eq:recurisive_osy_u}) and (\ref{eq:recurisive_osy_f_B}). 
 
 Once we  know how to compute the surviving network sizes $f_{A_{2i+1}}$ and $f_{B_{2i+2}}$ at each stage, the
 propagation of failures between the two networks is seen to be governed via (\ref{eq:recurisive_osy_p_A}) and (\ref{eq:recurisive_osy_p_B}) that reveal how the key quantities  $p_{A{2i+1}}$ and $p_{B_{2i+2}}$ used in computing $f_{A_{2i+1}}$ and $f_{B_{2i+2}}$, respectively, need to be updated based on the result of the last cascade stage.
  Collecting, a thorough analysis that reveals a  full understanding of the system behavior and robustness during the failure process is presented in equations (\ref{eq:recurisive_osy_p_A})-(\ref{eq:recurisive_osy_f_B}).

\section{Numerical Results}
\label{sec:numerical}
In this section, we confirm our analytic results through numerical simulations under a wide range of parameter choices, with a particular focus on checking the accuracy of the results when the network size $N$ is finite.

For physical networks carrying a certain flow (i.e., network $A$ in our analysis), we consider different combinations of probability distributions for the load and free-space variables. Throughout, we consider three commonly used families of distributions: i) Uniform, ii) Pareto, and iii) Weibull. These distributions are chosen here because they cover a wide range of commonly used and representative cases. In particular,  uniform distribution provides an intuitive baseline. Distributions belonging to the Pareto  family are also known as a {\em power-law} distributions and have been observed in many real-world networks including the Internet, the citation network, as well as power systems \cite{pahwa2010topological}. 
Weibull distribution is widely used in engineering problems involving reliability and survival analysis, and contains several classical distributions as special cases; e.g., Exponential, Rayleigh, and Dirac-delta.

The corresponding probability density functions are defined below for a generic random variable $L$.
\begin{itemize}
\item 
Uniform Distribution: $L \sim U(L_{\textrm{min}},L_{\textrm{max}})$. The density is given by
\[
p_L(x)=\frac{1}{ L_{\textrm{max}}  - L_{\textrm{min}} } \cdot \1{ L_{\textrm{min}} \leq x  \leq L_{\textrm{max}} }
\]
\item Pareto Distribution: $L \sim {Pareto}(L_{\textrm{min}}, b)$. With
$L_{\textrm{min}}>0$ and $b>0$, the density is given by
\[
p_L(x) =  L_{\textrm{min}}^{b} b x^{-b-1} \1{x \geq L_{\textrm{min}}}.
\]
To ensure that $\bE{L}=b L_{\textrm{min}}/(b-1)$ is finite, we also enforce $b>1$. Distributions belonging to the Pareto  family are also known as a {\em power-law} distributions and have been extensively used in many fields including power systems.
\item 
Weibull Distribution: $L \sim Weibull(L_{\textrm{min}},\lambda, k)$. With $\lambda, k, L_{\textrm{min}}>0$, the density is given by
\[
p_L(x) = \frac{k}{\lambda} \left(\frac{x-L_{\textrm{min}}}{\lambda} \right)^{k-1} e^{-\left(\frac{x-L_{\textrm{min}}}{\lambda} \right)^{k}}
 \1{x \geq L_{\textrm{min}}}.
 \]
The case $k=1$ corresponds to the exponential distribution, and $k=2$ corresponds to Rayleigh distribution. The mean is given by 
$\bE{L}=L_{\textrm{min}}+ \lambda \Gamma (1+1/k)$, where $\Gamma(\cdot)$ is the gamma-function given by
$\Gamma(x)=\int_{0}^{\infty} t^{x-1} e^{-t} dt$. 
\end{itemize}

As explained in Section \ref{subsec:detailed_model},
the cyber-network where a node is only functional when it belongs to the giant component (i.e., network $B$ in our analysis) is generated according to the configuration model  with degree distribution $\{d_k\}_{k=0}^{\infty}$. In the simulations, we consider two representative cases given below:
\begin{itemize}
\item
 Erd\H{o}s-R\'{e}nyi (ER) network model \cite{erdos1959random,erdos1960evolution,bollobas1998random}.
 This corresponds to having the degree distribution $d_k$ follow a Binomial distribution, i.e., $d_k \sim \textrm{Binomial}(N-1; \frac{\langle d \rangle}{N-1})$; as before  $\langle d \rangle$ gives the mean node degree.
 \item The scale-free (SF) network model  \cite{BarabasiAlbert}.
We consider the case where the degree distribution $\{d_k\}_{k=0}^{\infty}$ is a {\em power-law} with {\em exponential cut-off}, 
 which was observed \cite{ClausetShaliziNewman} in many real networks
including the Internet; i.e., we have
\begin{equation}
  d_k =
    \begin{cases}
      0 & \text{if $k=0$ }\\
    
      \frac{1}{\text{Li}_{\gamma}(e^{-1/\Gamma})}k^{-\gamma}e^{-k/\Gamma} & \text{if $k=1,2,\cdots$},
    \end{cases}       
    \label{eq:power_cut_off}
\end{equation}
where $\gamma$ is the power exponent, $\Gamma$ is the cut-off point, and $\text{Li}_m(z):=\sum_{k=1}^{k=\infty} z^k k^{-m}$ is the normalizing constant. 
 \end{itemize}
 
 We remind that although we restrict our attention to these special cases in the simulations, our analysis applies under more general  degree distributions as well. 

\subsection{Fiber Network Coupled with ER Network}
\label{sec:numerical_A}
The Erd\H{o}s-R\'{e}nyi graph is one of the most basic and widely used network models and often serve as a starting point in  simulations. In our  study, we start with $N$ nodes, and connect each pair of vertices with an edge
with probability $\langle d \rangle/(N-1)$ independently from each other.
When $N$ is large, this is equivalent to generating the network via the configuration model using a {\em Poisson} degree distribution with mean $\langle d \rangle$. 


\begin{figure}[!t]
\centering
{\includegraphics[width=0.5\textwidth]{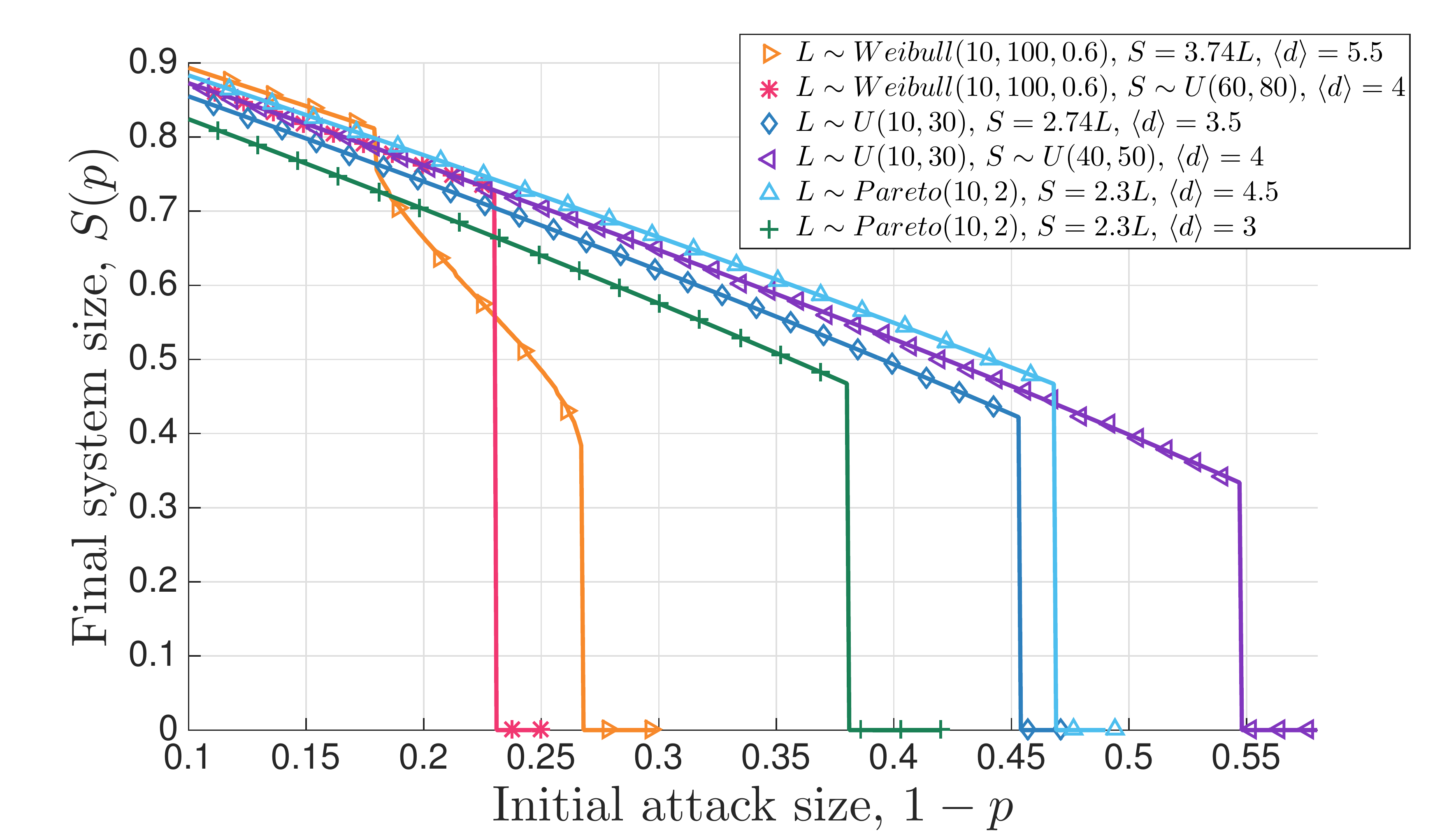}}
\caption{\sl Final system size under different network settings, including different load-free space distributions in the physical network and different mean degree in the cyber network. Analytic results are represented by lines, whereas simulation results are represented by symbols (averaged over 100 independent runs). We see that in each case theoretical results match the simulation results very well.}
\label{fig:general_match_addi}
\end{figure}

First, we confirm our main result presented in Sec. \ref{sec:results} concerning the final system size $S(p)$, i.e., the mean fraction of surviving nodes at the end of cascading failures initiated by a random attack that removes $1-p$ fraction of nodes in network $A$. In all simulations, we fix the number of nodes in both networks at $N=10^5$, and for each set of parameters being considered (i.e., the distribution $p_{LS}(x,y)$, the attack size $1-p$ in network $A$, and the mean degree $\langle d \rangle$ in network $B$), we run 100 independent experiments.  
The results are shown in Figure
\ref{fig:general_match_addi} where symbols represent the {\em empirical} value of the final system size $S(p)$ (obtained by averaging over 100 independent runs for each data point), and lines represent the
analytic results computed via (\ref{eq:recurisive_osy_p_A})--(\ref{eq:recurisive_osy_f_B}). We see that theoretical results match the simulations very well in all cases. This suggests that although asymptotic in nature, our main result can still be helpful when the network size $N$ is finite. The specific distributions used in Figure
\ref{fig:general_match_addi} are as follows: From left to right, we have 
i) in network $A$ (the physical network), $L$ is Weibull with $L_{\textrm{min}}=10,\lambda =100, k=0.6$ and $S=\alpha L$ with $\alpha = 3.74$; in network $B$ (the cyber network) the mean degree $\langle d \rangle = 5.5$;
ii) in network $A$, $L$ is Weibull with $L_{\textrm{min}}=10,\lambda =100, k=0.6$ and $S$ is Uniform over $[60,80]$; in network $B$ $\langle d \rangle = 4$;
iii) $L$ is Uniform over $[10,30]$ and and $S=\alpha L$ with $\alpha = 2.74$; in network $B$ $\langle d \rangle = 3.5$;
iv) $L$ is Uniform over $[10, 30]$ and $S$ is Uniform over $[40, 50]$; $\langle d \rangle = 4$;
v) $L$ is Pareto with $L_{\textrm{min}}=10$, $b=2$, $S=\alpha L$ with $\alpha=2.3$; $\langle d \rangle = 4.5$;
vi) $L$ is Pareto with $L_{\textrm{min}}=10$, $b=2$, $S=\alpha L$ with $\alpha=2.3$; $\langle d \rangle = 3$.

The plots in Figure \ref{fig:general_match_addi} show how different load-free space distributions in network $A$ as well as the mean degree in network $B$  affect the system behavior. For example, with the mean degree of network $B$ is fixed to $\langle d \rangle = 4$, the two different cases considered in Figure \ref{fig:general_match_addi}, one where the initial loads in network $A$ follow a Weibull distribution (magenta asterisk) and the other where the initial loads follow a Uniform distribution (purple triangle) lead to vastly different system behavior against attacks. When load in network $A$ follows Weibull distribution, the final system size drops to zero at a point where the attack size is around $0.23$, meaning that any random attack that kills more than $23\%$ of the nodes will destroy the entire system. On the other hand, if the load and free space follows Uniform distribution, the system is quite robust and can sustain initial attack sizes up to $0.55$ without collapsing. Similarly, when we fix the distribution in network $A$, we can see the effect of mean degree in the cyber network on system robustness: when initial load in physical network follows $\textrm{Pareto}(10,2)$ distribution, and free space is given by $S=\alpha L$ with $\alpha=2.3$, we see that increasing mean degree of network $B$ from $\langle d \rangle = 3$ (green cross) to $\langle d \rangle = 4.5$ (light blue triangle) leads to a substantial increase on the final system size at {\em all} attack sizes; i.e.,  the interdependent CPS becomes more robust. This is intuitive since higher $\langle d \rangle$ values lead to a cyber-network $B$ with higher levels of connectivity enabling the entire CPS to sustain larger attacks while maintaining a larger fraction of nodes in its giant component. 

An interesting observation from Figure \ref{fig:general_match_addi} is that in all cases, the final drop of the system size to zero takes place through a {\em first-order} (i.e., discontinuous) transition\footnote{The nomenclature concerning the order of transitions is adopted from the studies on phase transition in Physics; simply put, first (resp.~second) order transitions are associated with {\em discontinuous} (resp.~continuous) variations.}, 
 making it difficult to predict system behavior from previous data (in response to attacks with larger than previously observed size).
 In fact, this abrupt failure behavior is reminiscent of the real-world phenomena of unexpected large-scale system collapses; i.e., cases where seemingly identical attacks/failures leading to entirely different consequences. 
 We also see that our model can lead to a rich set of  behaviors to increasing attack sizes. For instance, when the initial load follows a Weibull distribution, depending on the parameters, it is possible to observe an abrupt first-order transition with no prior indication of system collapse at smaller attack sizes (magenta asterisk), as well as a first-then-second order transition (orange triangle) before the system size drops to zero through a final first-order transition. These behaviors are due to the intrinsic characters of different distributions, and should be considered in designing CPS where the physical network may be governed by different flow distribution types. 

\begin{figure}[!t]
 \vspace{-4mm}
\centering
{\includegraphics[width=0.5\textwidth]{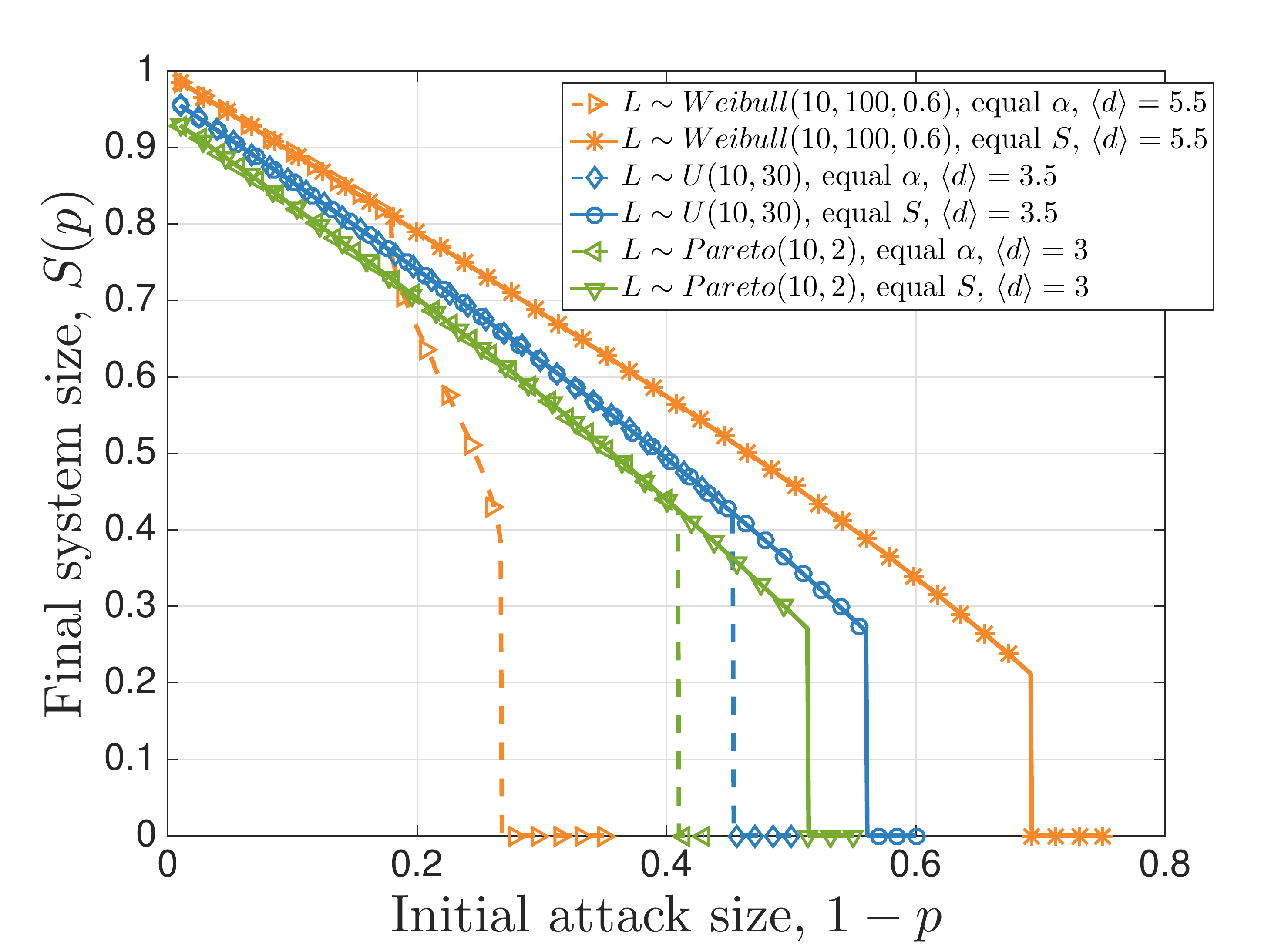}}
\caption{\sl Final system size under equal free space (solid lines with symbols) or equal tolerance factor (dashed lines with symbols) when network $B$ is a ER graph with fixed mean degree. The symbols are empirical results over 100 independent runs on network size $N=10^5$, and lines (dashed or solid) represent analytic results. We can see in all cases equal free space greatly improves system robustness by allowing the system to sustain a larger initial attack size and still not collapsing.
 \vspace{-4mm}
}
\label{fig:equal_S_ills}
\end{figure}

From a design perspective, it is of interest to understand how the robustness of the interdependent system 
can be improved or even maximized under certain constrains. To gain insights on this, we fix the mean degree in network $B$ (the cyber network), and explore the effect of the allocation (i.e., distribution) of node capacities in the physical network.  
A key determining factor of system robustness is expected to be the free-space distribution as it specifies the extra load a node can receive from the failed ones before it fails due to overloading. The vast majority of the literature and most real world applications employ a {\em linear} free-space allocation scheme where the free-space assigned to a node is set to be a fixed proportion of its initial load. In other words, it is assumed  that $S=\alpha L$, where $\alpha$ is the {\em tolerance factor} and is usually a fixed value \cite{MotterLai,WangChen,Mirzasoleiman,CrucittiLatora} used for the entire network. We already showed in \cite{zhang2016optimizing} that in a single flow-carrying network, allocating every node exactly the same free-space leads to a higher robustness (at any attack size $1-p$) than the commonly used setting of equal tolerance factor (with the comparison made when the total free-space in the entire network is fixed). In fact, in the single network case, the robustness is shown to be {\em maximized} when all nodes receive the same free space.

Our numerical simulations, presented in Figure \ref{fig:equal_S_ills}, shows that the above conclusion still applies in interdependent networks. Namely, assigning every node the same free space provides a much better overall system robustness as compared to the widely used setting of equal tolerance factor (i.e., linear free-space allocation). To provide an overall evaluation of the system robustness, we define the critical attack size $1-p^{\star}$ as the minimum attack size that breaks down the whole system. Thus, the larger $1-p^{\star}$ is, the more robust will  the system be since it can sustain larger attacks. 
In Figure \ref{fig:equal_S_ills}, the comparison between the equal free-space and equal tolerance factor allocations are made with the mean free space $\mathbb{E}[S]$ being fixed (i.e., the total free space in the network is constrained). We see that compared to the equal tolerance factor scheme, the equal free-space allocation enables the system to sustain much larger attacks. In fact, in the case of Weibull distribution, the robustness  is almost 2.5 times higher in the case of equal free space as compared to the case with equal tolerance factor; i.e., $1-p^{\star}=0.7$ vs. $1-p^{\star}=0.28$. 

\begin{figure*}[!t]
\centering
{\includegraphics[width=0.65\textwidth]{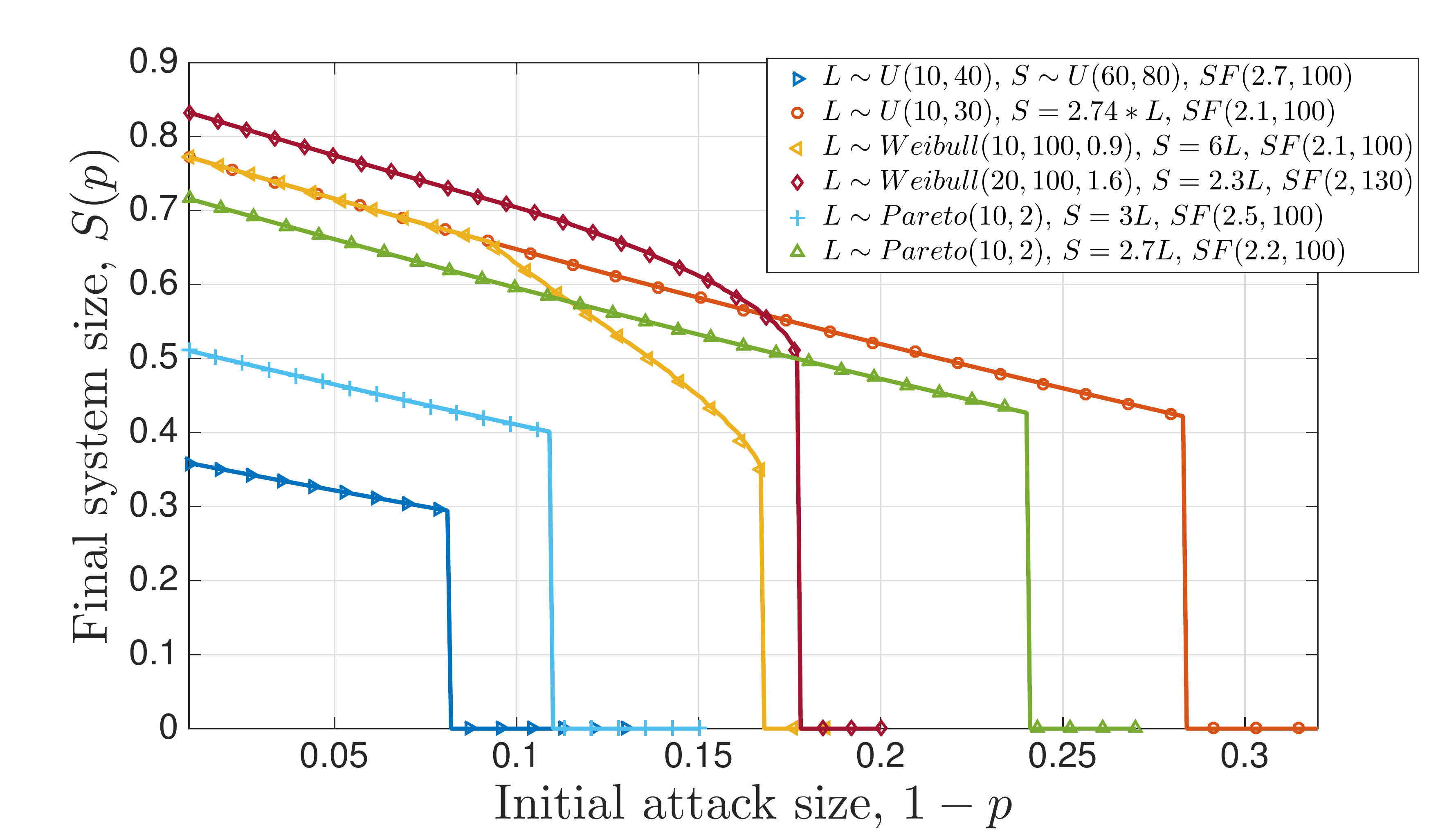}}
\caption{\sl Final system size under different network settings, including different load-free space distributions in the physical network and different exponent in the scale-free cyber network. Analytic results are represented by lines, whereas simulation results are represented by symbols (averaged over 100 independent runs). We see that in each case theoretical results match the simulation results very well.}
\label{fig:general_match_SF}
\vspace{-2.5mm}
\end{figure*}

\subsection{Fiber Network Coupled with SF Network}

Although the ER graph constitutes a simple and useful network model, networks in most real-world applications might have significantly different structure and robustness behavior against attacks. For instance, scale-free networks (SF model) were shown \cite{albert1999internet}  to exhibit fundamentally different robustness behavior with ER networks; the former is very robust against random attacks but fragile against {\em targeted} attacks, while  the situation is exactly   the opposite for the latter. In order to better understand the impact of the topology of the cyber-network on the overall robustness of an interdependent CPS, we consider in this section the case where the cyber network (network $B$) has a power-law degree distribution with exponential cutoff. In addition to being observed  in many real-world networks including the Internet \cite{ClausetShaliziNewman}, power-law distributions with exponential cut-off also ensure that all moments of the node degree are {\em finite}, which helps certain convergences take place faster (i.e., with smaller $N$).



\begin{figure*}
\begin{multicols}{2}
    \includegraphics[width=.98\linewidth]{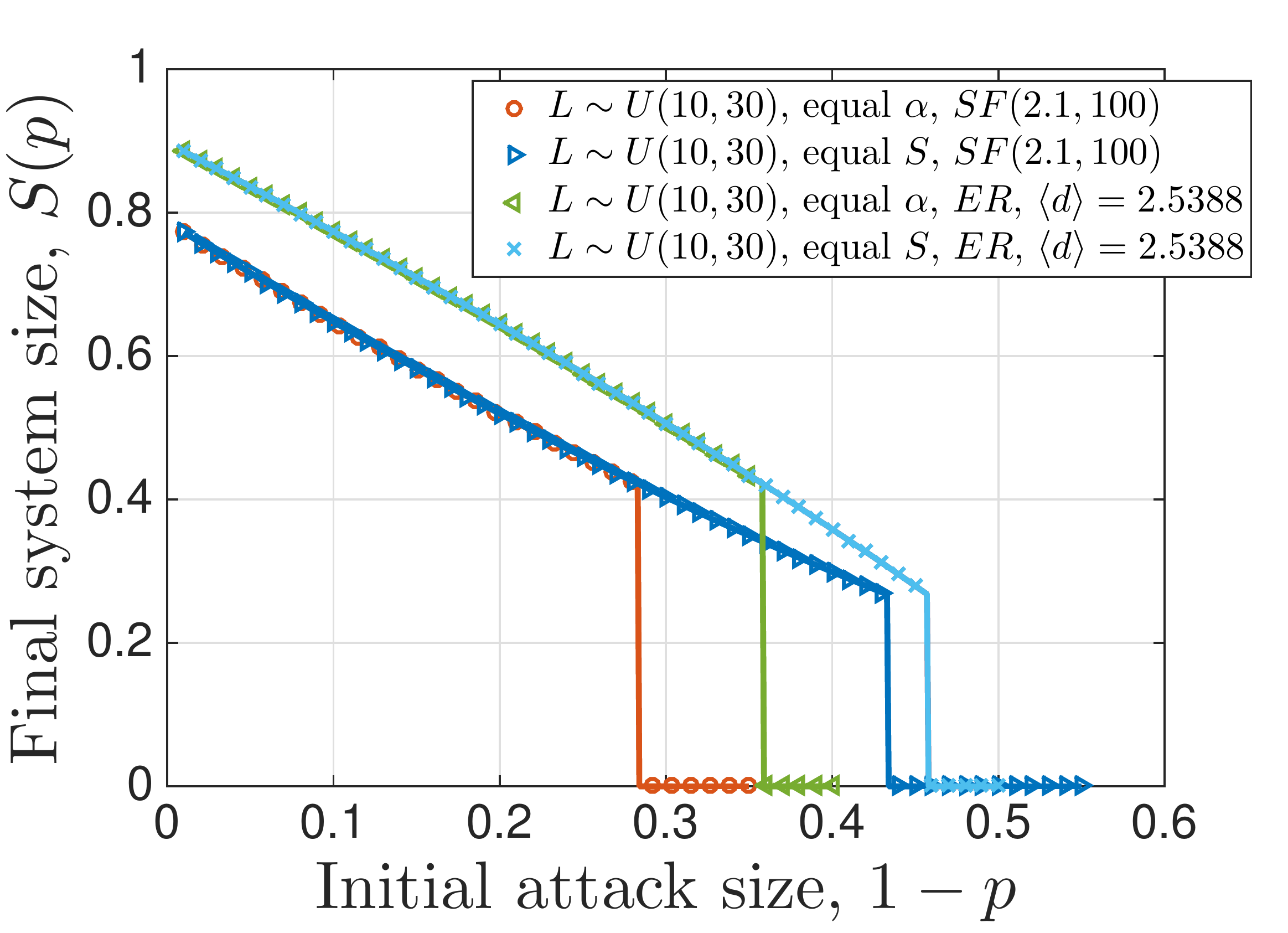}\par 
    \includegraphics[width=.98\linewidth]{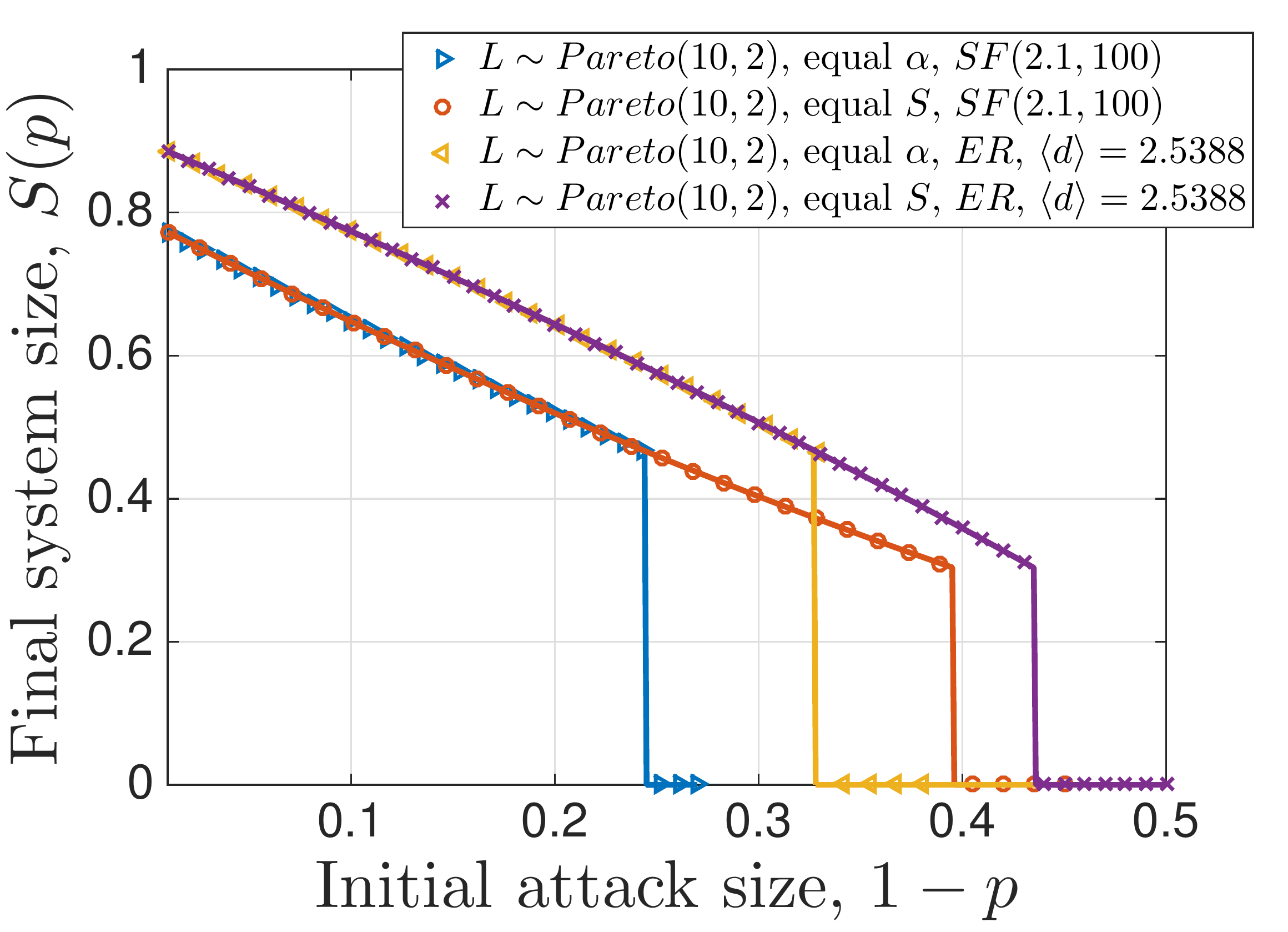}\par 
    \end{multicols}
\begin{multicols}{2}
    \includegraphics[width=.98\linewidth]{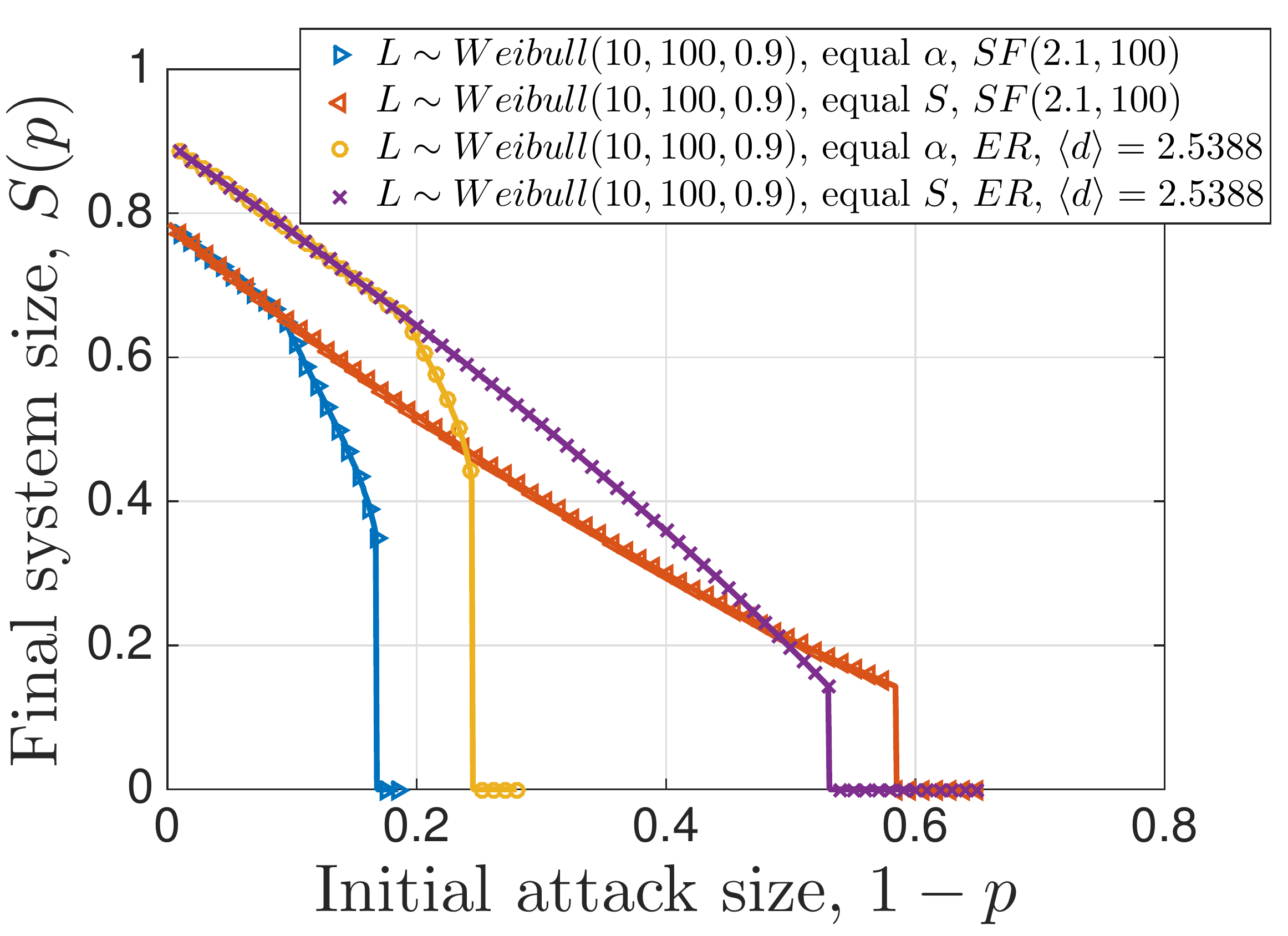}\par
    \includegraphics[width=.98\linewidth]{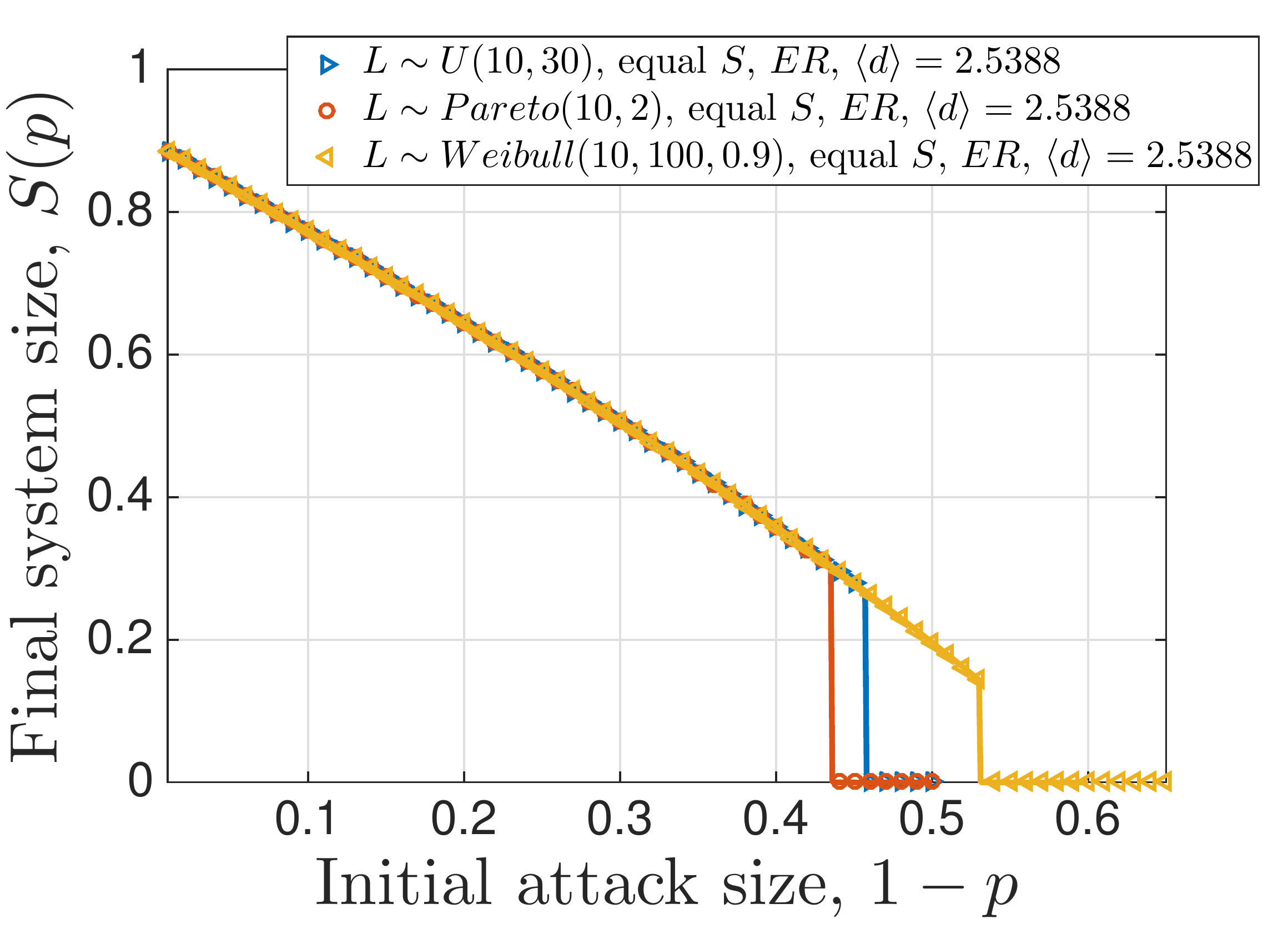}\par
\end{multicols}
\vspace{-3mm}
\caption{ 
Comparison of final system size when equal tolerance factor (equal $\alpha$) and equal free space (equal $S$) schemes are used. The mean value of free space is kept the same, as well as the mean degree in SF and ER networks. In all cases, equal $S$ outperform the widely used equal $\alpha$ scheme. The effect of topology in the cyber network is not unitary:  in some cases ER leads to better robustness, while in other cases SF is better, contradicting the results \cite{BarabasiAlbert} concerning the robustness of single networks.}
\label{fig:equal_S_alpha_compare}
\end{figure*}

In Figure \ref{fig:general_match_SF}, we verify our analytic results when network $B$ (cyber network) has a degree distribution in the form of a power-law with exponential cut-off;  this is denoted by $SF(\gamma,\Gamma)$, where 
$\gamma$  is the power exponent and  $\Gamma$ is the cut-off parameter given at (\ref{eq:power_cut_off}). 
In all cases, we fix the number of nodes in both networks to be $N=10^6$, and 
consider several different load-\lq free space' distributions in network $A$ and different $(\gamma, \Gamma)$ values for network $B$ (while noting that in real-world networks, it is often observed that $2<\gamma<3$).
The simulation results are obtained by averaging over 100 independent experiments for each data point and it is seen that 
they are in very good agreement with the analytic results.

Next, we seek to obtain an overall understanding of how the free-space allocation in the physical network together with the topology of the cyber network (ER vs.~SF model) affect the system robustness. With the discussion from Section \ref{sec:numerical_A} in mind, we consider the widely used equal tolerance factor allocation (equal $
\alpha$), where the free-space $S$ is a fixed factor $\alpha$ of the load on a node (i.e., $S=\alpha*L$) and the equal free-space allocation scheme (equal $S$) that was shown \cite{zhang2016optimizing} to be optimal in a single physical network. For fairness, all comparisons are made under the same initial load distribution, and with the mean free-space in network $A$ and the mean node degree in network $B$ being fixed. 

The results are shown in Figure \ref{fig:equal_S_alpha_compare}. We can see that no matter how the initial load is distributed, i.e., whether it's Uniform, Pareto or Weibull, and despite of the structure of the cyber network being SF network or ER network, equal free-space allocation can greatly improve system robustness as compared to the equal tolerance factor allocation. We observe that when all nodes in network $A$ are given the same free-space, the overall interdependent CPS can sustain a much larger initial attack size without collapsing; i.e., it has a much larger critical attack size. For example in the case of Weibull distributed load with SF network, the system can only take around 16.8\% of initial attack size when using equal $\alpha$, but can sustain a initial attack that removes 58\% of the nodes when equal $S$ is used, making the system about 3 times more robust in terms of the critical attack size. 

We also see in Figure \ref{fig:equal_S_alpha_compare} that the topology of the cyber network affects the robustness 
of the interdependent CPS in an intricate way, with some cases showing the exact opposite of what would have been expected from the results on single networks. In particular, SF networks are known \cite{BarabasiAlbert} to be more robust than ER networks against random attacks. This is often attributed to the fact that SF networks typically have a few nodes with very high degrees and the network will likely contain a large connected component unless these high-degree nodes are removed (which is unlikely to happen if the attack is {\em random}); this dependence on a few nodes is exactly what makes SF networks very fragile against a {\em targeted} attack.
In the case of the interdependent CPS model, we see that the comparison of the overall robustness between the cases where they cyber network is SF or ER  is a much more complicated matter. In fact, depending on 
the load-\lq free-space' distribution in the physical network, the cyber network being SF does {\em not} always lead to a better robustness than the case with ER. For example, in the upper two plots in Figure \ref{fig:equal_S_alpha_compare} where the initial load is Uniform and Pareto, respectively, the cases with the ER network leads to a better robustness than that with SF. In the bottom left picture where the initial load in the physical networks is Weibull, the situation is even more intricate. With equal $\alpha$, the case where the cyber-network is ER leads to a better robustness, while SF network performs better (in terms of the critical attack size) under the equal-$S$ allocation. This shows that an integrated CPS can not be designed in the most robust way by considering the physical and cyber counterparts separately. Instead, a holistic design approach is needed where the robustness of the CPS as a whole is considered.  



\section{Conclusion}
\label{sec:conclusion}
We have studied the robustness of an interdependent system against cascading failures initiated by a random attack. This is done through a novel model where the constituent networks exhibit inherently different intra-dependency characteristics. In particular, inspired by many applications of interdependent cyber-physical systems (CPSs), our model consists of a flow network where failure of a node leads to flow redistribution and possible further failures due to {\em overloading} (i.e., the flow on a node exceeding its capacity), and a cyber-network where nodes need to be a part of the largest connected cluster to be functional. We derive relations for the dynamics of cascading failures, characterizing the mean fraction of surviving nodes from each network at every stage of the cascade. This leads to deriving the mean fraction of nodes that ultimately survive the cascade as a function of the initial attack size. Through numerical simulations, we confirm our analysis and derive useful insights concerning the robustness of interdependent CPSs. 

There are many open directions for future work. First of all, the simplistic one-to-one interdependence model used here can be replaced by more sophisticated and realistic dependency models. A good starting point would be to consider a model where every node is assigned $k$ inter-links and can continue to function as long as at least one of its $k$ support nodes in the other network is functional. It would be interesting to study the trade-off between the number of inter-links and the resulting improvements in overall system robustness; one might also consider a heterogeneous allocation of inter-links and study the optimal (in the sense of maximizing robustness) way to assign inter-links subject to certain constraints \cite{YaganQianZhangCochranLong}. It would also be interesting the consider more complicated flow redistribution models based on network topology, rather than the equal redistribution model considered here. Finally, it would be interesting to study the system robustness under targeted attacks (where the set of nodes to be attacked is chosen carefully by an adversary) rather than the case of random attacks considered in this paper.

\bibliographystyle{abbrv}
\bibliography{sample,references,refs2}

\appendix
\setcounter{equation}{0}
\setcounter{section}{0}
\renewcommand{\thesection}{\Alph{section}}
\renewcommand{\theequation}{\thesubsection.\arabic{equation}}
\renewcommand{\thetheorem}{\thesubsection.\arabic{theorem}}

 \section*{Proof of the Main Result}

We now prove the main result of the paper, i.e., Theorem \ref{thm:main}.
This entails recursively deriving the {\em mean} fraction of surviving nodes from both networks at each stage $t=1, 2, \ldots$ of the cascade process. The cascade process starts at time $t=0$ with a random attack that kills $1-p$ fraction of the nodes from network $A$. As mentioned earlier, we assume an asynchronous model where at stages $t=1, 3, \ldots$ we consider the failures in network $A$ and in stages $t=2, 4, \ldots$ we consider the failures in network $B$. In this manner, we keep track of the subset of vertices $A_1 \supset A_3 \supset \ldots \supset A_{2i+1}$ and
$B_2 \supset B_4 \supset \ldots \supset B_{2i}$ that represent the functioning (i.e., surviving) nodes at the corresponding stage of the cascade. We let $f_{A_{i}}$ denote the {\em relative} size of the surviving set of nodes from network $A$ at stage $i$, i.e.,
\[
f_{A_{i}} = \frac{|A_i|}{N}, \qquad i = 1, 3, 5, \ldots
\]
We define $f_{B_i}$ similarly as
\[
f_{B_{i}} = \frac{|B_i|}{N}, \qquad i = 2, 4, 6, \ldots
\]

 \subsection{Computing the Functional Component in Physical Network} \label{sec:func_phy}

First, we consider the intra-failures within the physical network $A$ and derive $f_{A_{i}}$ for 
$i=1, 3, \ldots$. Initially, $1 - p_{A_1}$ fraction of nodes are attacked (or, failed) randomly in network $A$, where $p_{A_1} \in [0,1]$. The flow in the failed nodes will get redistributed (equally) to all remaining  nodes that are not attacked. Each such node will now have an increased load on them. If the extra load received is greater than the free-space on a node (equivalently, if the current load is greater than their capacity), it will fail resulting in another round of redistribution, and so on and so forth.

 In our previous work 
 \cite{zhang2016optimizing}, we analyzed the cascading failures in a single flow network and obtained results that allows computing i) the fraction of nodes that will survive at the steady-state of cascades; and ii) the extra load that each surviving line will be carrying, once the {\em steady-state} is reached. 
 Here, this result, presented below as Theorem 
 \ref{thm:intra_A_old_result} for convenience, will be used in calculating the outcome of each round of intra-failures in network $A$. 
 \begin{theorem}[\cite{zhang2016optimizing}]
{\sl Consider a flow network of $N$ nodes  
as described in Section \ref{subsec:detailed_model}. Let
$L$ and $S$ denote generic random variables following the same distribution (i.e., $p_{LS}(x,y)$) with initial loads $L_1,\ldots,L_N$,
and free spaces $S_1, \ldots, S_N$, respectively.
Let $n_{\infty}(p)$ be the {\em mean} fraction   of nodes that are still functioning at the steady of cascading failures initiated by a random attack on $(1-p)$-fraction of the nodes. Then, we have
\begin{equation}
n_{\infty}(p) = p \bP{S > x^{\star}}.
\label{eq:final_size}
\end{equation}
with $x^{\star}$ denoting the {\em extra load carried by each node that survives at the steady-state} and is given by
\begin{align}
&x^{\star} = \min\bigg\{x \in (0, \infty]: \label{eq:main_condition}
\\
& \qquad \qquad \qquad \qquad \bP{S > x} \left( x + \bE{L ~|~ S > x} \right) \geq \frac{\bE{L}}{p}\bigg\}
\nonumber
\end{align}
 }
 \label{thm:intra_A_old_result}
 \end{theorem}

In words, this result states that when $(1-p)$-fraction of nodes fail randomly in a flow network where the initial load-\lq free-space' distribution is given by $p_{LS}$, the cascading failures (caused by recursive load redistribution) will stop with $p \mathbb{P}[S> x^{\star}]$-fraction of nodes surviving, and the   load-\lq free-space' distribution of surviving nodes 
will be updated as 
\begin{align}
L^{(\cdot)} &~\sim  L+x^{\star}~|~S>x^{\star}
\label{eq:updated_load_distribution}\\
S^{(\cdot)} &~\sim S-x^{\star}~|~S>x^{\star}
\label{eq:updated_space_distribution}
\end{align}
where $x^{\star}$ is as defined in (\ref{eq:main_condition}). These observations as well as 
(\ref{eq:final_size}) follow immediately from the fact that $x^{\star}$ gives the amount of extra load each surviving node carries at the steady-state; see \cite{zhang2016optimizing} for a detailed explanation of this fact. For example, (\ref{eq:final_size}) is seen easily as the probability that a node is {\em not} in the initial set of $(1-p)$-fraction of failed nodes, {\em and} has more free-space than the $x^{\star}$ amount of load it receives through redistribution from nodes that failed at any step of the cascade. Also, given that a node survives this cascade of failures, the probability distribution of its load-\lq free-space' values need to be updated via  conditioning on the fact that its free space satisfies 
$S>x^{\star}$. Noting also that each surviving node will now carry $x^{\star}$ amount of {\em extra} load in addition to its initial load, and will have $x^{\star}$ amount of {\em less} free-space than it started with, we obtain
(\ref{eq:updated_load_distribution})
and
(\ref{eq:updated_space_distribution}).

In the forthcoming discussion we will repeatedly use the fact that the set of nodes that survive at the steady-state still form a flow network as described in Section \ref{subsec:detailed_model}, with its fractional size reduced to $p\mathbb{P}[S>x^{\star}]$ and its load-\lq free-space' distribution updated via (\ref{eq:updated_load_distribution})
-
(\ref{eq:updated_space_distribution}). 
For the interdependent network model under consideration,
 the intra-failures in network $A$ will be followed by failures in network $B$ first due to inter-dependence, and then due to intra-dependence in $B$, with the latter possibly triggering further failures in $A$; e.g., see Figure \ref{fig:failure_process}). 
 For instance, for any stage $i=0,1,\ldots$, let us assume that the intra-failures in network $A$ has just resulted in $A_{2i-1}$. This will be followed by the aforementioned failures in $B$ leading to $B_{2i}$. 
  Since intra-failure dynamics of network $B$ is completely {\em independent} from network $A$, the impact of the failures in $B$ to network $A$ will be   will be equivalent to a {\em random} attack launched on $A_{2i-1}$; in fact, it will be equivalent to a random attack targeting $1-|B_{2i}|/|A_{2i-1}|$-fraction of the nodes. The discussion above shows that we can  compute the outcome of this random attack again from Theorem \ref{thm:intra_A_old_result} and obtain the size of $A_{2i+1}$, and so on.


 With these in mind, we now start by computing $f_{A_1}$
 which gives the resulting size of network $A$ after a random attack to $1-p_{A_1}$-fraction of nodes; i.e., 
 $f_{A_1}$ gives the fraction of nodes in $A_1$.
 Using Theorem \ref{thm:intra_A_old_result}, we immediately get 
 \begin{equation} \label{eq:step_1_a}
     f_{A_1} = p_{A_1}\bP{S > x^{\star}_1}
 \end{equation}
 where $x^{\star}_1$ is given from
 \begin{align} \label{eq:step_1_b}
&x^{\star}_1 = \min\bigg\{x \in (0, \infty]: 
\\
&  \qquad \qquad \qquad \bP{S > x} \left( x + \bE{L ~|~ S > x} \right) \geq \frac{\bE{L}}{p_{A_1}}\bigg\}
\nonumber
\end{align}
We also know from Theorem \ref{thm:intra_A_old_result} that the $f_1$ fraction of nodes that constitute the sub-network $A_1$
 have their loads and free-spaces coming from the updated distributions 
 \begin{align}
L^{(1)} &~\sim  L+x^{\star}_1~|~S>x^{\star}_1
\label{eq:L_1}
\\
S^{(1)} &~\sim S-x^{\star}_1~|~S>x^{\star}_1
\label{eq:S_1}
\end{align}

We find it useful to denote by $Q_i$ the extra load that each surviving node carries at stage $i=1,3,5, \ldots$; i.e., $Q_i$ denotes the extra load each node in $A_i$ carries. This leads to $Q_1 = x^{\star}_1$, and we get
\begin{equation} \label{eq:step_1_a_new}
     f_{A_1} = p_{A_1}\bP{S > Q_1}
 \end{equation}
where
\begin{align} \label{eq:step_1_b_new}
&Q_1 = \min\bigg\{x \in (0, \infty]: 
\\
&  \qquad \qquad \qquad \bP{S > x} \left( x + \bE{L ~|~ S > x} \right) \geq \frac{\bE{L}}{p_{A_1}}\bigg\}
\nonumber
\end{align}
Setting $f_{B_0}=p$, $f_{A_{-1}}=1$, and $Q_{-1}=0$ for convenience, 
these establish the desired results (\ref{eq:recurisive_osy_p_A})-(\ref{eq:recurisive_osy_f_A}) for $i=0$. 
 
 After the intra-failures in network $A$ stabilizes, we check the effect of these on network $B$, again according to the asynchronous cascading failure model summarized in Figure \ref{fig:failure_process}.
 Due to the one-to-one interdependence between network $A$ and $B$, when cascading failures stop at network $A$, network $B$ reduces to $\bar B_2$ which has the same size with $A_1$; in fact, $\bar B_2 = \{b_j : a_j \in A_1\}$.
 According to our model, the intra-failures in the cyber-network $B$ are governed by the {\em giant-component} rule, i.e., at each stage a node in $B$ is functional if and only if it belongs to its largest connected component. This will lead to a reduction of
 $\bar B_2$ to its largest connected component, defined as
 $B_2$. Due to one-to-one interdependence model, we will next observe $|\bar B_2-B_2|$ nodes failing from network $A$, more precisely from $A_1$.
 It should be clear that the nodes that belong to $|\bar B_2-B_2|$ have failed solely according to the initial intra-topology of network $B$ which is constructed independently from network $A$. Therefore, the removal of  $|\bar B_2-B_2|$ nodes from $A_1$ will be equivalent to a {\em random} attack launched on $A_1$ that removes
 \[
  \frac{|\bar B_2-B_2|}{|A_1|} 
=\frac{|A_1|-|B_2|}{|A_1|} 
 = 1- \frac{f_{B_2}}{f_{A_1}}
 \]
fraction of nodes.

 As in (\ref{eq:recurisive_osy_p_A}) with $i=1$, we define $p_{A_3}=\frac{f_{B_2}}{f_{A_1}}$. Now, the impact of intra-failures in network $A$ can be computed once again from Theorem \ref{thm:intra_A_old_result}. This time, a random attack of size $1-p_{A_3}$ will be launched on $A_1$ whose fractional size is given by $f_{A_1}$, and the distribution of load and free-space of its nodes are as given in 
 (\ref{eq:L_1})-(\ref{eq:S_1}). From Theorem \ref{thm:intra_A_old_result}, we get
 \begin{equation} \label{eq:step_2_a}
     f_{A_3} = f_{A_1}p_{A_3}\bP{S^{(1)}>x^{\star}_3}
 \end{equation}
 where $x^{\star}_3$ is given from
 \begin{align} \label{eq:step_2_b}
&x^{\star}_3 = \min\bigg\{x \in (0, \infty]: 
\\
&    \qquad \bP{S^{(1)} > x} \left( x + \bE{L^{(1)} ~|~ S^{(1)} > x} \right) \geq \frac{\bE{L^{(1)}}}{p_{A_3}}\bigg\}
\nonumber
\end{align}
Also, each surviving node in $A_3$ will now have an updated load and free-space distribution, given as  
  \begin{align}
L^{(3)} &~\sim  L^{(1)}+x^{\star}_3~|~S^{(1)}>x^{\star}_3
\nonumber
\\
S^{(3)} &~\sim S^{(1)}-x^{\star}_3~|~S^{(1)}>x^{\star}_3.
\nonumber
\end{align}
 From (\ref{eq:L_1}) and (\ref{eq:S_1}), this is equivalent to 
   \begin{align}
L^{(3)} &~\sim  L+Q_3~|~S>Q_3
\label{eq:L_3}
\\
S^{(3)} &~\sim S-Q_3~|~S>Q_3,
\label{eq:S_3}
\end{align}
as we set $Q_3 = x^{\star}_1 + x^{\star}_3$. As before, $Q_3$ denotes the extra load that each node in $A_3$ carries.

 We now use  (\ref{eq:L_1})-(\ref{eq:S_1}) and definitions of $Q_1,Q_3$ to simplify (\ref{eq:step_2_a}) and (\ref{eq:step_2_b}). First, observe that
 \begin{align}\label{eq:simp_b}
     \bP{S^{(1)}>x^{\star}_3} &= \bP{S -x^{\star}_1 > x^{\star}_3 ~|~ S >x^{\star}_1} 
     \nonumber \\
     &= \bP{S>Q_3 ~|~ S>Q_1} 
     \nonumber
 \end{align}
Similarly,
\begin{align}
&\bP{S^{(1)} > x} \left( x + \bE{L^{(1)} ~|~ S^{(1)} > x} \right)
\nonumber
\\
& = \bP{S > x^{\star}_1+x ~|~S >x^{\star}_1} 
\left( x + \bE{x^{\star}_1 +L ~|~ S > x^{\star}_1} \right)
\nonumber \\
& = \frac{\bP{S > Q_1+x}}{\bP{S >Q_1}}
\left( x + Q_1+ \bE{L ~|~ S > Q_1} \right)
\nonumber
\end{align}
since $x > 0$. 
Finally, we have
 \begin{equation} \label{eq:simp_a}
 \bE{L^{(1)}} = \bE{L  +x^{\star}_1~|~ S  > x^{\star}_1} = Q_1 + \bE{L~|~ S > Q_1} 
 \nonumber
 \end{equation}
Reporting these into  (\ref{eq:step_2_a}) and (\ref{eq:step_2_b}), we now get
 \begin{equation} \label{eq:step_2_a_new}
     f_{A_3} = f_{A_1}p_{A_3}\bP{S>Q_3 ~|~ S>Q_1}
 \end{equation}
 where
 \begin{align}\label{eq:step_2_b_new}
    &Q_3 = Q_1 + 
     \min\bigg\{x \in (0, \infty]: \frac{\bP{S > Q_1+x}}{\bP{S >Q_1}}\times
\\
&    
\quad \times \left( x + Q_1+ \bE{L ~|~ S > Q_1} \right) \geq \frac{Q_1 + \bE{L~|~ S > Q_1} }{p_{A_3}}\bigg\}
\nonumber
\end{align} 
The desired results (\ref{eq:recurisive_osy_p_A})-(\ref{eq:recurisive_osy_f_A}) of
Theorem \ref{thm:main} are now established for $i=1$.  
 
The cascade process will continue in the same manner, by considering the failures in network $B$. Again, the impact of 
intra-failures in $B$ will be equivalent to a random attack launched at $A_3$ that fail
\[
  \frac{|\bar B_4-B_4|}{|A_3|} 
=\frac{|A_3|-|B_4|}{|A_3|} 
 = 1- \frac{f_{B_4}}{f_{A_3}} := 1-p_{A_5}
 \]
 fraction of nodes. We can use Theorem \ref{thm:intra_A_old_result}
 again to compute the fraction of nodes from $A_3$ that survive this random attack. This time we should note that the fraction of nodes in $A_3$ equals $f_3$, while the load and free-space values of nodes in $A_3$ follow (\ref{eq:L_3}) and (\ref{eq:S_3}), respectively. This gives us 
 \begin{equation} \label{eq:step_3_a}
     f_{A_5} = f_{A_3}p_{A_5}\bP{S^{(3)}>x^{\star}_5}
 \end{equation}
 where $x^{\star}_3$ is given from
 \begin{align} \label{eq:step_3_b}
&x^{\star}_5 = \min\bigg\{x \in (0, \infty]: 
\\
&    \qquad \bP{S^{(3)} > x} \left( x + \bE{L^{(3)} ~|~ S^{(3)} > x} \right) \geq \frac{\bE{L^{(3)}}}{p_{A_5}}\bigg\}
\nonumber
\end{align}

Once again, we can simplify (\ref{eq:step_3_a}) and (\ref{eq:step_3_b}) using (\ref{eq:L_3}), (\ref{eq:L_3}) and setting $Q_5=Q_3 + x^{\star}_5$. We omit the details here in the interest of brevity, but it is straightforward to see that
\begin{equation} \label{eq:step_3_a_new}
     f_{A_5} = f_{A_3}p_{A_5}\bP{S>Q_5 ~|~ S>Q_3}
 \end{equation}
 where
 \begin{align}\label{eq:step_3_b_new}
    &Q_5 = Q_3 + 
     \min\bigg\{x \in (0, \infty]: \frac{\bP{S > Q_3+x}}{\bP{S >Q_3}}\times
\\
&    
 \times \left( x + Q_3+ \bE{L ~|~ S > Q_3} \right) \geq \frac{Q_3 + \bE{L~|~ S > Q_3} }{p_{A_5}}\bigg\}
\nonumber
\end{align} 
The results (\ref{eq:recurisive_osy_p_A})-(\ref{eq:recurisive_osy_f_A}) are now established for $i=2$.

The form of the recursive equations concerning the functional size of network $A$ is now clear: We have 
(\ref{eq:recurisive_osy_p_A})-(\ref{eq:recurisive_osy_f_A})
for each $i=0,1, \ldots$ upon setting
 \begin{equation}
     Q_{2i+1} = \sum_{k=1}^{i} x^{\star}_{2k+1} = Q_{2i-1} + x^{\star}_{2i+1}.
 \end{equation}
 As already mentioned, $Q_{2i+1}$ denotes the extra load that
 each node in $A_{2i+1}$ carries. 
 Using  (\ref{eq:recurisive_osy_p_A})-(\ref{eq:recurisive_osy_f_A}), we can compute the size, $f_{A_{2i+1}}$ of the functioning nodes in the physical network $A$ at any stage $i$ of the cascading failure process. Next, we will look at how we can compute the size $f_{B_{2i}}$ of functioning component in the cyber network $B$.

 \subsection{Computing the Giant Component in Cyber Network} \label{sec:compute_gc}

We now explain how the remaining fraction of functional nodes in the cyber-network $B$ can be computed at each stage of the cascading failures; i.e., we will establish (\ref{eq:recurisive_osy_p_B}), (\ref{eq:recurisive_osy_u}), and (\ref{eq:recurisive_osy_f_B}). The main idea we will use in this proof is the fact that the failures that take place in network $A$ 
at each cascade step due to flow redistribution will be seen as a {\em random} attack to the remaining portion of network $B$. This follows from the intra-topology of network $B$ being independent from the load and free-space values of the corresponding nodes in $A$. 

With this in mind, the following result established in \cite{newman2002spread,MolloyReed}
will be used repeatedly. 
\begin{theorem}
\label{thm:intra_B_old}
{\sl Consider a network $B$ generated randomly according to the configuration model with degree distribution $\{d_k\}_{k=0}^{\infty}$; i.e., we have $\bP{\textrm{degree of node $b_i$} = k} = d_k$ for each $k=0,1, \ldots$ and $i=1,\ldots, N$.
The mean degree is denoted by $\langle d \rangle$, i.e., we have $\langle d \rangle=\sum_{k=0}^{\infty} k d_{k}$.
Let $f_B(p)$ be the {\em mean} fraction of nodes that are still functioning after a random attack is launched on $(1-p)$-fraction of the nodes in $B$. In other words, with $\bar{B}$ denoting the network obtained by deleting every node of $B$ independently with probability $(1-p)$, let $f_B(p)$ give the fraction of nodes that are in the giant component of $\bar{B}$.
We have
\begin{align}
    f_{B}(p) & = p \left( 1 - \sum_{k=0}^{\infty} d_k \left(1- u^{\star} p\right)^k \right) \label{eq:osy_f_B}
\end{align}
where
\begin{align}
\hspace{-3mm}     u^{\star} = \max \left\{\hspace{-1mm}u \in [0,1]\hspace{-.5mm}: u= 1\hspace{-.5mm}-\hspace{-.5mm}\sum_{k=0}^{\infty} \frac{k d_k}{\langle d \rangle}  (1-u p)^{k-1}  \right\} \label{eq:osy_u} 
\end{align}
}
\end{theorem}

Theorem \ref{thm:intra_B_old} can be used directly to derive (\ref{eq:recurisive_osy_p_B})-(\ref{eq:recurisive_osy_f_B}) for $i=0$, i.e., to compute the fraction of nodes in $B_2$. 
As stated before, the initial attack on $1-p$ fraction of the nodes and subsequent failures in network $A$ due to flow redistribution leads to network $A$ reducing to $A_1$ whose relative size is given by $f_{A_1}$. Due to the one-to-one interdependence between networks $A$ and $B$, this will lead to network $B$ reducing to $\bar B_2$ which has the same size with $A_1$; in fact, $\bar B_2 = \{b_j : a_j \in A_1\}$. In light of the aforementioned independence between the intra-topology of network $B$ and the failures that lead $A$ to reduce to $A_1$, the reduction of $B$ to $\bar B_2$ will be equivalent to a random attack launched on $B$ that kills
\[
1-|\bar B_2| = 1-f_{A_1}:= p_{B_2}
\]
fraction of the nodes. Thus, the size of the giant component of $\bar B_2$ (denoted 
as $B_2$ with relative size $f_{B_2}$)
can be computed from  Theorem \ref{thm:intra_B_old} by replacing $p$ with $p_{B_2}$. This yields
\begin{align}\nonumber
    f_{B_2} & = p_{B_2} \left( 1 - \sum_{k=0}^{\infty} d_k \left(1- u_2 p_{B_2}\right)^k \right)
\end{align}
where
\begin{align}\nonumber
     u_2 &= \max \left\{u \in [0,1]: u= 1-\sum_{k=0}^{\infty} \frac{k d_k}{\langle d \rangle}  (1-u p_{B_2})^{k-1}  \right\}. 
\end{align}
These establish (\ref{eq:recurisive_osy_p_B})-(\ref{eq:recurisive_osy_f_B}) for $i=0$.

For the subsequent stages, we use the following result that was introduced in \cite{Buldyrev}. Let network $B$ be generated as described in Theorem \ref{thm:intra_B_old}. Assume that, as in the case studied here, $B$ goes through a series of random attacks, where after each attack only the largest connected component of the non-attacked part remains functional. For example, let $B$ go through the following process. 
\begin{align}
\nonumber   & B \xrightarrow[(1-p_{2})-\textrm{fraction}]{\textrm{random attack}} \bar B_2 \xrightarrow[\textrm{comp.}]{
\textrm{giant}} B_2 \xrightarrow[1-p_{4}]{\textrm{attack}} \bar B_4 \xrightarrow[\textrm{comp.}]{
\textrm{giant}} B_4 
\\ \label{eq:equiv_process1}
& ~~~~ \rightarrow \cdots \rightarrow B_{2i} \xrightarrow[1-p_{2i+2}]{\textrm{attack}} \bar B_{2i+2} \xrightarrow[\textrm{comp.}]{
\textrm{giant}} B_{2i+2}
\end{align}
Since all the attacks here are random, this process is equivalent (in terms of the fractional size of $B_{2i+2}$) to the case where a single random attack (with appropriate size to be discussed next) is applied to $B$ followed by the computation of the giant component of the remaining network. In other words, the process described in (\ref{eq:equiv_process1}) yields statistically the same fraction of nodes in $B_{2i+2}$ with the process 
\begin{align}   & B \xrightarrow[(1-p_{2}\cdot p_{4} \cdots p_{{2i+2}})-\textrm{fraction}]{\textrm{random attack}} \bar B \xrightarrow[\textrm{comp.}]{
\textrm{giant}} B_{2i+2}
\label{eq:equiv_process2}
\end{align}

This result can be understood by considering the point of view of a single fixed node, say $v_b$ in $B$, and calculating its probability of being in $B_{2i+2}$. Given that the network is attacked sequentially with probabilities of failing any node being $1-p_{2}$, $1-p_{4}, \ldots 1-p_{{2i+2}}$, the probability for $v_b$ to {\em not} fail in any of these stages is given by 
\begin{align}
p_{2} \cdot p_{4} \cdots p_{2i+2}.
\label{eq:multiplicative_p_B}
\end{align}
This probability is the same with the case if $B$ experiences only one attack that kills 
$(1-p_2\cdot p_{4} \cdots p_{2i+2})$-fraction of nodes. Secondly, $v_b$ will survive this process and will be in $B_{2i+2}$ if it is also in the giant component of the remaining part of $B$. We know from \cite{MolloyReed} that $B$ can have at most one component whose fractional size is positive, with all other components having size $o(n)$. This is also true for all the intermediary networks $B_2, B_4, \ldots$. Thus, the process of reducing the network to its giant component can be done only once after all the attacks have been applied, without affecting the fractional size of the resulting giant component; i.e., the {\em small} components (with size $o(n)$) eliminated at each stage in (\ref{eq:equiv_process1}) will have no affect on the fractional size of $B_{2i+2}$. 

With this result, we can now establish (\ref{eq:recurisive_osy_p_B})-(\ref{eq:recurisive_osy_f_B}) for all $i=1, 2, \ldots$. 
For each $i$, the fractional size $f_{B_{2i+2}}$ of $B_{2i+2}$ can be computed from Theorem \ref{thm:intra_B_old} in view of the equivalence of the processes (\ref{eq:equiv_process1})
and (\ref{eq:equiv_process2}). In particular, $f_{B_{2i+2}}$ is calculated from 
(\ref{eq:osy_f_B})-(\ref{eq:osy_u}) as the fractional size of the giant component of $\bar B$, obtained by deleting every node of $B$ with probability $1-p_{B_{2i+2}}$. Here, $p_{B_{2i+2}}$ is the probability for any node in $B$ to not have been attacked in any of the previous stages, and can be computed as in (\ref{eq:multiplicative_p_B}). In particular, it is given by the multiplication of the probabilities of a node $v_b$ {\em not} losing its support in network $A$ in consecutive stages. We can compute this probability recursively as 
\begin{align}
\nonumber
p_{B_{2i+2}} &= p_{B_{2i}} \bP{\begin{array}{c}
\textrm{$v_b$ does not lose its support}       \\
      \textrm{in $A$ at stage $2i+1$} 
\end{array}}
\\ \label{eq:final_step_p_B}
&= p_{B_{2i}} \frac{|A_{2i+1}|}{|B_{2i}|}
\\
&= p_{B_{2i}} \frac{f_{A_{2i+1}}}{f_{B_{2i}}}
\end{align}
where (\ref{eq:final_step_p_B}) follows from the independence of failures in $A$ with the topology of network $B$. We now established (\ref{eq:recurisive_osy_p_B}) for each $i=0,1, \ldots$, and (\ref{eq:recurisive_osy_u})-(\ref{eq:recurisive_osy_f_B}) follow from Theorem \ref{thm:intra_B_old} and the discussion that follows.

\end{document}